\documentclass{article}

\usepackage[numbers]{natbib}
\usepackage{algorithmic}
\usepackage{algorithm}
\usepackage{array}
\usepackage[caption=false,font=normalsize,labelfont=sf,textfont=sf]{subfig}
\usepackage{textcomp}
\usepackage{stfloats}
\usepackage{newunicodechar}
\newunicodechar{，}{,}
\usepackage{verbatim}
\usepackage{booktabs}
\usepackage{multirow}
\usepackage{amsmath}
\usepackage{tabularx}

\usepackage{PRIMEarxiv}

\usepackage[utf8]{inputenc} 
\usepackage[T1]{fontenc}    
\usepackage{hyperref}       
\usepackage{url}            
\usepackage{booktabs}       
\usepackage{amsfonts}       
\usepackage{nicefrac}       
\usepackage{microtype}      
\usepackage{lipsum}
\usepackage{fancyhdr}       
\usepackage{graphicx}       
\graphicspath{{media/}}     

\title{MALF: A Multi-Agent LLM Framework for Intelligent Fuzzing of Industrial Control Protocols
}

\author{
  Bowei Ning \\
  School of Artificial Intelligence \\
  Shenyang University of technology \\
  Shenyang\\
  \texttt{2020183@stu.syuct.edu.cn} \\
   \And
  Xuejun Zong, Kan He \\
  School of Information Engineering \\
  Shenyang University of Chemical Technology \\
  Shenyang\\
  \texttt{xuejun\_zong@syuct.edu.cn} \\
}

\begin{document}
\maketitle

\begin{abstract}
Industrial control systems (ICS) are vital to modern infrastructure but increasingly vulnerable to cybersecurity threats, particularly through weaknesses in their communication protocols. This paper presents MALF (Multi-Agent LLM Fuzzing Framework), an advanced fuzzing solution that integrates large language models (LLMs) with multi-agent coordination to identify vulnerabilities in industrial control protocols (ICPs). By leveraging Retrieval-Augmented Generation (RAG) for domain-specific knowledge and QLoRA fine-tuning for protocol-aware input generation, MALF enhances fuzz testing precision and adaptability.
The multi-agent framework optimizes seed generation, mutation strategies, and feedback-driven refinement, leading to improved vulnerability discovery. Experiments on protocols like Modbus/TCP, S7Comm, and Ethernet/IP demonstrate that MALF surpasses traditional methods, achieving a test case pass rate (TCPR) of 88–92\% and generating more exception triggers (ETN). MALF also maintains over 90\% seed coverage and Shannon entropy values between 4.2 and 4.6 bits, ensuring diverse, protocol-compliant mutations.
Deployed in a real-world Industrial Attack-Defense Range for power plants, MALF identified critical vulnerabilities, including three zero-day flaws, one confirmed and registered by CNVD. These results validate MALF's effectiveness in real-world fuzzing applications.
This research highlights the transformative potential of multi-agent LLMs in ICS cybersecurity, offering a scalable, automated framework that sets a new standard for vulnerability discovery and strengthens critical infrastructure security against emerging threats.
\end{abstract}


\section{Introduction}

The rapid evolution of industrial control systems (ICS) has shifted their architecture from the traditional ISA-95 hierarchical model to more dynamic, networked cloud-edge-end frameworks. This transformation, driven by advancements in information and communication technology, facilitates real-time and reliable interconnection among production elements, enhancing operational efficiency\cite{ref1}. Key industrial control protocols (ICPs) enable cross-layer interoperability, supporting technologies like Time-Sensitive Networking and 5G\cite{ref2}. However, these protocols introduce unique security risks due to their reliance on deterministic, stateful communication patterns and vendor-specific implementations. For instance, Modbus/TCP lacks native encryption, making it susceptible to man-in-the-middle attacks, while OPC UA’s complex metadata structures create parsing vulnerabilities exploitable via malformed packets. Such weaknesses, when exploited, can disrupt safety-critical processes (e.g., PLC command injection) or enable lateral movement in ICS networks \cite{ref3}.

Fuzz testing, a pivotal dynamic vulnerability discovery technique, involves generating and injecting malformed inputs to expose latent weaknesses\cite{ref4}.  Fuzz testing of ICPs presents distinct challenges compared to general-purpose fuzzing\cite{ref5}. The structured and context-dependent nature of ICPs demands a comprehensive understanding of their message sequences, timing constraints, and communication contexts\cite{ref6}. Moreover, testing in these environments must account for concurrency and timing sensitivities that are critical to security assessments\cite{ref7}. These requirements necessitate specialized methods to address the intricate characteristics of ICPs effectively.For example, Modbus/TCP employ highly structured headers and payloads, requiring in-depth protocol knowledge to produce meaningful message seeds\cite{ref8}. Inefficient mutation processes stemming from inadequate input diversity can undermine fuzzing efforts. 

The integration of LLMs and artificial intelligence (AI) offers promising solutions to these challenges. LLMs, with their advanced contextual understanding and natural language processing abilities, are well-suited for generating context-aware inputs for ICPs\cite{ref9}. By leveraging multi-agent systems, LLMs can enable modular collaboration, with agents specializing in input generation, vulnerability analysis, and robustness assessment. This approach simulates human-like teamwork\cite{ref10}, optimizing fuzz testing in the complex, modularized environments of modern ICS. Such a strategy enhances testing efficiency, uncovers deeper vulnerabilities, and overcomes the limitations of traditional fuzzers.

This work presents a novel framework that leverages multi-agent LLMs for fully automated, modular, and intelligent fuzz testing of ICPs. The proposed framework addresses the distinct challenges posed by modern industrial environments, such as flattened network architectures and diverse communication protocols, through adaptive input generation and iterative learning. By dynamically evolving the fuzzing process, it identifies increasingly sophisticated vulnerabilities over time. Additionally, new metrics are introduced to evaluate fuzzing effectiveness, emphasizing both the depth of security analysis and protocol coverage. The main contributions are summarized as follows:

\begin{itemize}
\item{We introduce MALF, a fully automated fuzzing framework specifically designed for ICPs. MALF addresses key ICS fuzzing challenges such as protocol compliance, mutation diversity, and adaptive feedback optimization, without requiring manual intervention.}
\item{We employ advanced LLM methods, including RAG and QLoRA fine-tuning, to dynamically generate protocol-aware test cases. This enhances fuzzing processes such as seed generation, mutation, and real-time feedback adjustments, improving testing precision and adaptability.}
\item{We establish a real-world benchmark platform for fuzz testing within an industrial attack-defense range. MALF outperforms existing fuzzing methods, demonstrating superior efficiency and test case diversity. It also uncovers critical vulnerabilities, including latent flaws in ICPs, thereby extending fuzz testing to the industrial internet ecosystem.}
\end{itemize}

\section{Related Works}
The field of fuzz testing has evolved considerably over the past few decades, driven by advancements in methodologies and technologies. Recent developments have expanded traditional fuzzing techniques to address the complexities of modern systems. This section provides an overview of recent progress in fuzz testing, focusing on mutation- and generation-based fuzzing, advanced protocol fuzzing, and the integration of multi-agent systems powered by LLMs.

\subsection{Traditional Mutation- and Generation-Based Fuzzing}
Early fuzz testing primarily relied on mutation- and generation-based techniques. Mutation-based fuzzers modify a set of valid inputs to create diverse, potentially invalid or unexpected inputs, aiming to trigger vulnerabilities or unforeseen behaviors\cite{ref11}. Prominent examples include AFL (American Fuzzy Lop)\cite{ref12} and libFuzzer\cite{ref13}, which have demonstrated notable success, particularly in file-based and application fuzzing. For instance, AFL uses a coverage-guided mutation strategy that iteratively refines the input corpus, enhancing the discovery of new code paths and vulnerabilities\cite{ref14}. This method has proven effective in various applications, providing valuable insights into bugs and security flaws.
In contrast, generation-based fuzzing creates inputs from scratch based on predefined grammars or rule sets\cite{ref15}. This approach is particularly effective for structured data formats, where input structure must be precisely controlled. Tools like Peach Fuzzer\cite{ref16} and Sulley\cite{ref17} focus on generating inputs that adhere to specific formats, making them well-suited for testing systems that rely on structured communication protocols or parsers. However, traditional fuzzing techniques encounter significant challenges when applied to complex, stateful protocols common in modern industrial environments\cite{ref18}. These methods often lack the ability to understand protocol semantics and struggle to explore expansive input state spaces effectively.

\subsection{Advanced Protocol Fuzzing Techniques}
As industrial and networking systems adopt increasingly complex communication protocols, fuzzing techniques have evolved to address their intricacies. Protocol fuzzing specifically targets vulnerabilities in protocol implementations\cite{ref19}, ensuring robustness against attacks.
Recent advances include state-aware fuzzing, which accounts for protocol statefulness to improve test accuracy. Tools such as AFLNet\cite{ref20}, Boofuzz\cite{ref21}, and Nsfuzz\cite{ref22} extend traditional approaches by incorporating state tracking. For instance, AFLNet employs a state-machine model to better understand protocol states, enabling targeted testing of stateful systems like HTTP, FTP, and proprietary ICPs. This approach is particularly effective in uncovering vulnerabilities tied to specific interaction sequences.
Machine learning (ML) has further enhanced protocol fuzzing by predicting critical states and guiding fuzzing engines toward unexplored areas\cite{ref23}. Tools like NEUZZ\cite{ref24} use neural networks to optimize mutation strategies, increasing code coverage and bug detection rates. However, ML-based methods face limitations with heterogeneous, opaque ICPs, which demand deeper contextual understanding\cite{ref25}.

\subsection{LLM-Driven Fuzzing Techniques}
The advent of LLMs has introduced a transformative approach to fuzz testing, particularly for addressing challenges in complex protocol fuzzing\cite{ref11}. LLM-driven frameworks, such as TitanFuzz\cite{ref26} and FuzzGPT\cite{ref27}, leverage the contextual awareness and generative capabilities of LLMs to create adaptive and intelligent fuzzing solutions.
These frameworks use multi-agent architectures to dynamically generate inputs, analyze responses, and refine fuzzing strategies based on real-time feedback. For example, TitanFuzz integrates Codex for protocol-specific input generation with agents that perform static and dynamic analysis, enabling nuanced exploration of protocol state spaces. This approach excels in environments where traditional methods struggle, such as ICPs with diverse communication stacks and intricate state transitions.
A key advantage of LLM-driven fuzzing lies in its ability to generate context-aware test cases that incorporate protocol history, intended functionality, and semantic constraints. This goes beyond deterministic, rule-based fuzzing, offering a more sophisticated means of uncovering vulnerabilities in complex systems\cite{ref28}. Furthermore, these frameworks include dynamic evaluation mechanisms that assess the functional correctness and security relevance of generated inputs, adapting strategies to maximize coverage and vulnerability detection\cite{ref29}.

The evolution of fuzz testing reflects a progression from traditional mutation- and generation-based techniques to advanced protocol fuzzing and LLM-driven frameworks. Traditional methods are effective for simpler systems but often fail with the complexity and statefulness of modern ICPs. Advanced protocol fuzzing introduces state modeling and ML-based optimization, while LLM-driven systems leverage AI to provide adaptability, contextual awareness, and multi-agent collaboration. These innovations set the stage for more effective security analysis of ICS, ensuring resilience in increasingly interconnected environments.

\section{Motivation}
The increasing complexity of industrial control systems and their associated network protocols has made security testing a critical but highly challenging task\cite{ref30}. Traditional fuzzing techniques, whether based on mutation or generation, face significant limitations in dealing with the intricacies of modern industrial network protocols. These protocols often involve highly structured and layered communication mechanisms, require strict adherence to communication sequences, and integrate real-time constraints. As industrial systems evolve towards open, collaborative, and intelligent configurations, the demand for sophisticated, automated, and flexible fuzzing methods has become more urgent. This section delves into the motivation behind employing multi-agent LLM-driven approaches to address these challenges. 

\subsection{The Need for LLM-Based Fuzzing}
Recent advances in LLMs have demonstrated their ability to comprehend and generate contextually rich and structured data across various domains. Applying LLMs to the task of fuzzing industrial network protocols brings a promising opportunity to enhance both the diversity and quality of generated test cases. LLMs are capable of understanding and generating protocol-specific message formats\cite{ref31}, complex command sequences, and adherence to communication standards, which makes them an ideal candidate for testing structured communication environments such as industrial protocols. Furthermore, ICPs are highly structured and require deep contextual knowledge to create valid messages\cite{ref32}. LLMs, pre-trained on a vast amount of technical and structured data, provide a foundation for understanding these protocols, which can subsequently be refined to generate protocol-specific fuzzing inputs effectively. Utilizing LLMs provides an advantage in constructing highly varied and contextually aware seeds, which improves the potential of uncovering vulnerabilities that might be missed by traditional fuzzing methods. 

\subsection{Threat Model for Industrial Network Protocols} 
Industrial network protocols are subject to unique security challenges that differentiate them from conventional IT protocols. These protocols often run in environments that are highly sensitive to disruptions, such as critical manufacturing lines, power grids, and petrochemical facilities. An attacker that exploits a vulnerability in an industrial network protocol can cause significant disruptions, leading to financial losses, safety hazards, or even catastrophic failures. The threat model for industrial network protocols includes various attack vectors, such as unauthorized command injection, denial of service (DoS), and replay attacks\cite{ref33}. These threats exploit inherent weaknesses in how protocols manage authentication, command execution, and state management. Unlike typical IT network protocols, industrial communication protocols often lack robust security mechanisms, making them particularly vulnerable to fuzzing attacks that could identify flaws in their implementation. The consequences of these vulnerabilities are severe, ranging from equipment malfunction to complete operational downtime. 

\subsection{Limitations of Off-the-Shelf LLMs} 
While LLMs are inherently powerful, their application in ICPs fuzzing is not without challenges. Off-the-shelf LLMs, even those trained on a diverse range of internet-based technical texts\cite{ref34}, often lack the specialized knowledge required to understand the specificities of industrial network protocols. These models may generate syntactically correct but semantically incorrect messages that fail to capture the exact structure and operational requirements of the protocols, rendering the fuzzing process inefficient or incomplete. Furthermore, industrial protocols have intricate dependencies and specific input constraints that off-the-shelf LLMs struggle to fully comprehend without proper guidance or fine-tuning. For instance, LLMs may be unaware of the strict sequence in which certain commands must be executed or the contextual relationship between different protocol layers. This lack of domain-specific fine-tuning reduces their effectiveness when employed directly for fuzzing tasks in industrial environments. Another limitation lies in the LLMs' inability to autonomously evaluate the generated messages' validity within the protocol's context\cite{ref35}. Traditional LLM-based generation lacks a feedback mechanism that can iteratively refine the generated messages to ensure that they align with protocol-specific requirements. This limits the capability of achieving deep and meaningful testing coverage, especially for complex scenarios involving multiple interconnected devices and dynamic environments. 

\subsection{The Motivation for a Multi-Agent LLM Approach}
Given the challenges associated with off-the-shelf LLMs and the complexities of industrial protocols, we propose a multi-agent LLM-driven framework for industrial network protocol fuzzing. The idea is to leverage multiple specialized agents, each with distinct responsibilities, to collaboratively achieve comprehensive fuzzing outcomes. This approach allows us to modularize the fuzzing process into discrete tasks, such as protocol analysis, seed generation, mutation, and evaluation, assigning each task to an LLM-driven agent. The motivation for employing a multi-agent framework stems from the need for a distributed and parallel approach to tackle different aspects of the fuzzing process, allowing for greater coverage and efficiency. In this context, the agents operate in a feedback loop, where one agent generates protocol-compliant seeds, another performs mutations while ensuring syntactical and semantic correctness, and yet another evaluates the efficacy of these messages by assessing their impact on the target system. Such a collaborative, multi-agent setup aims to ensure that each generated message adheres to protocol-specific constraints while maximizing the diversity and potential of the test cases. Our work aims to build an intelligent, automated, and modular protocol fuzzing framework that incorporates the strengths of LLMs while overcoming their limitations through specialization and collaboration. This multi-agent architecture not only enhances the generation and validation of test inputs but also facilitates iterative learning and adaptation throughout the fuzzing process. By designing specialized agents with distinct yet complementary roles, we can significantly improve the efficiency of fuzzing, leading to more robust testing of industrial network protocols. The following sections will detail the architecture of our proposed multi-agent LLM fuzzing framework, including the operational dynamics of each module and the collaborative workflow that drives the entire system.

\section{Methodology}
\begin{figure*}[!b]
\centering
\includegraphics[width=5 in]{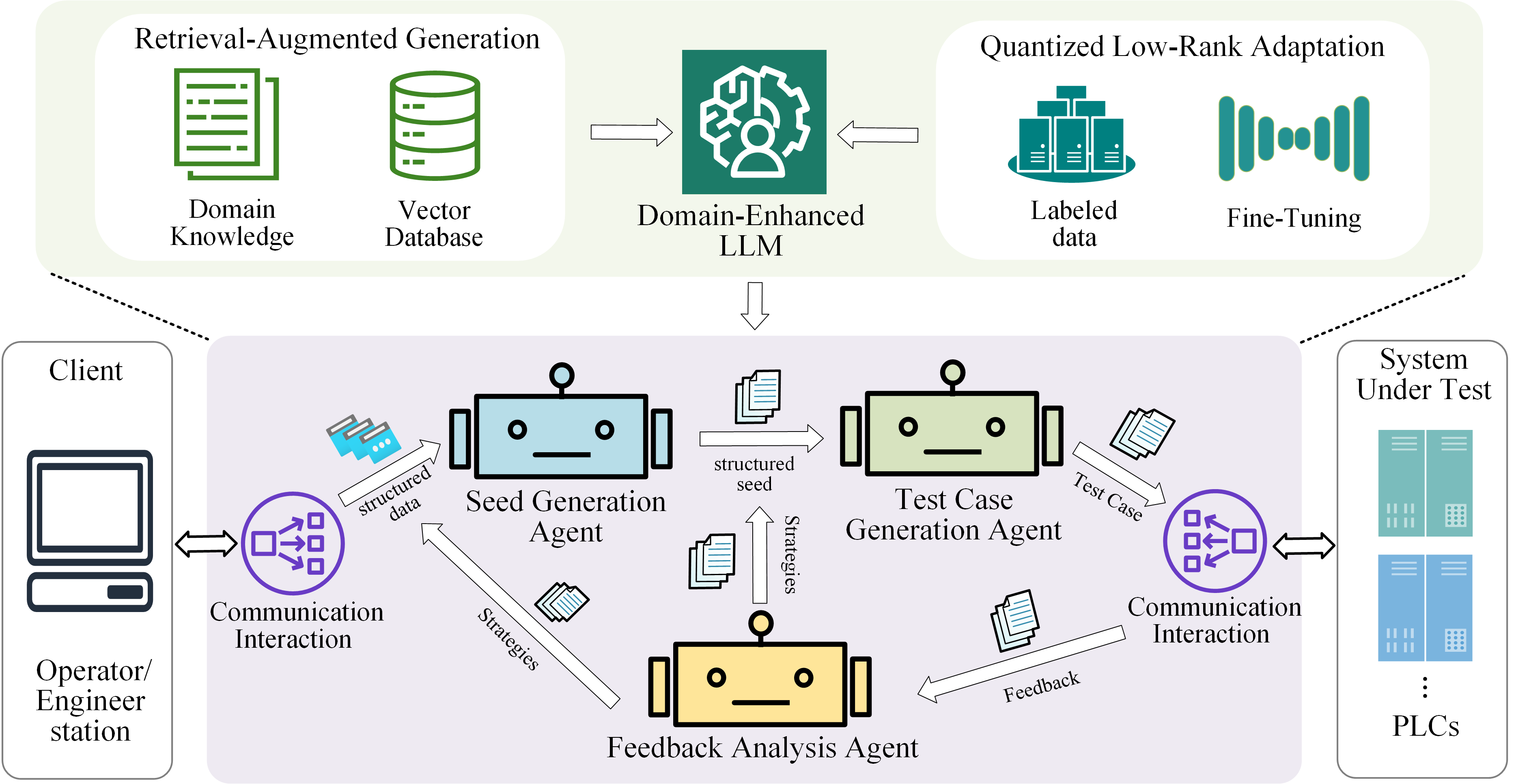}
\caption{The framework architecture of MALF.}
\label{fig_1}
\end{figure*}

Current fuzz testing methods struggle with complex ICPs due to challenges in protocol compliance and high-quality input generation. While many existing approaches use LLMs or deep learning for assistance, they often rely on manual intervention and fail to fully automate the fuzz testing loop. To overcome these limitations, we propose a multi-agent LLM-driven fuzzing framework (MALF) that automates the entire fuzz testing process for ICPs. As shown in Fig. \ref{fig_1}, MALF integrates four core components—Seed Generation Agent, Test Case Generation Agent, Feedback Analysis Agent, and Communication Interaction Module—to generate diverse test inputs, optimize testing strategies, and refine vulnerability detection in real time.

\subsection{Technical Foundations}
\begin{figure*}[!h]
\centering
\includegraphics[width=5in]{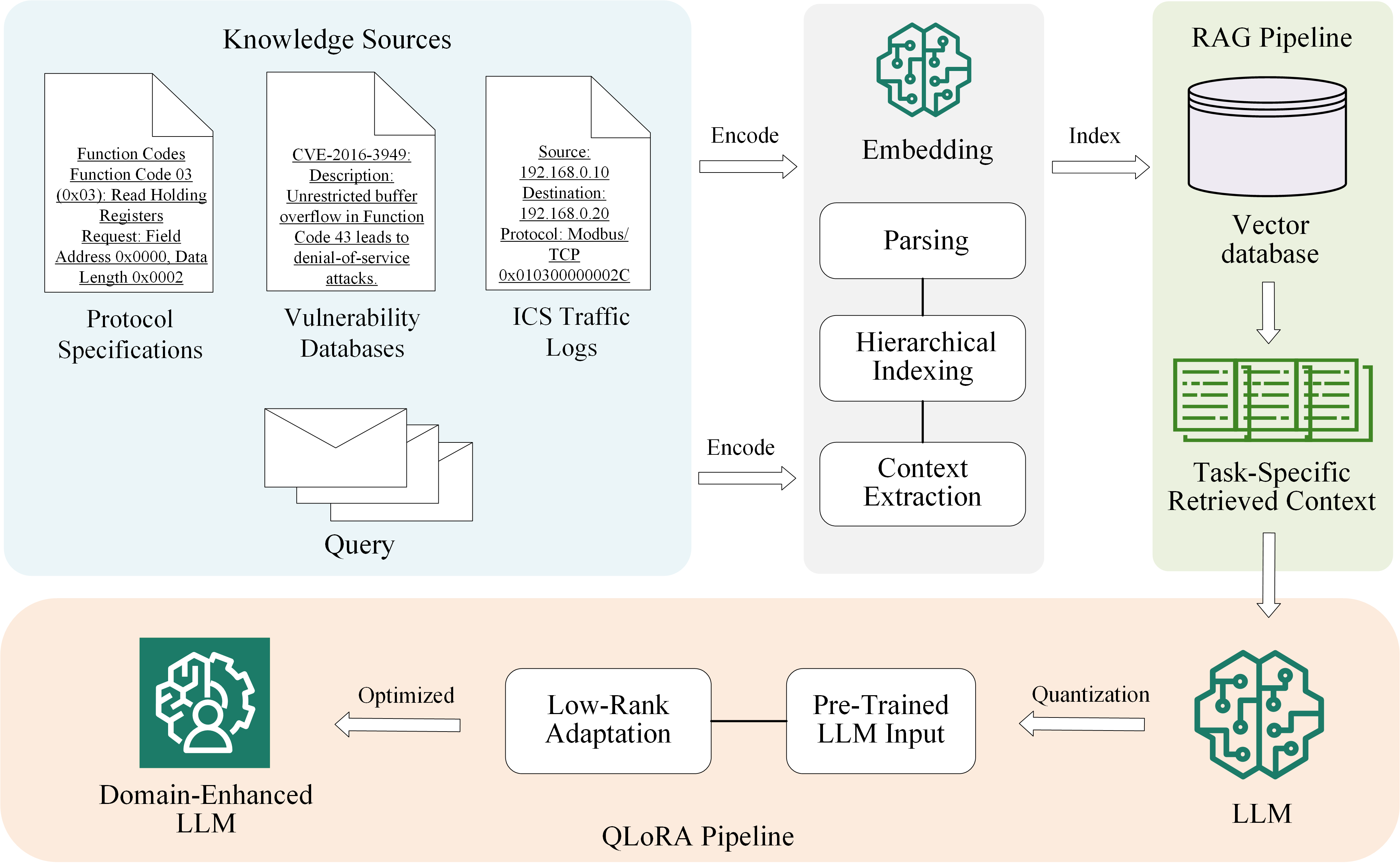}
\caption{Technical Architecture of the Multi-Agent LLM: RAG and QLoRA Pipelines}
\label{fig_2}
\end{figure*}

This section outlines the key techniques driving the MALF: RAG and QLoRA, enabling the integration of domain-specific knowledge into pre-trained LLMs for ICP fuzz testing. The technical architecture is shown in Fig. \ref{fig_2}.

\subsubsection{Retrieval-Augmented Generation}

RAG follows a retrieve-then-read approach to augment LLMs with protocol-specific knowledge, addressing their context size limitations\cite{ref36}. In MALF, RAG retrieves information (e.g., command structures, field formats, vendor-specific constraints) from structured ICP documents (e.g., Modbus/TCP, S7, Ethernet/IP) and vulnerability databases (e.g., CVE, CNVD). This ensures the LLMs produce domain-aware outputs. Instead of retraining models, we leverage the structured organization of ICP documents, typically in PDF format, to extract meaningful knowledge. The extraction process consists of the following:

\romannumeral1.\ Document Parsing and Indexing: We parse protocol documents to construct a hierarchical outline that captures section levels, titles, and page ranges. Regular expressions identify sections containing key terms, e.g., "function code," "data field," or "command format."

\romannumeral2.\ Content Extraction: For each identified section, subsections are recursively analyzed to extract detailed descriptions of specific commands or field attributes, e.g., “Function Code 03: Read Holding Registers.” These sections are consolidated into a domain knowledge base.

\romannumeral3.\  Knowledge Database: Across the analyzed protocol documents, we constructed a database comprising 320 pages and approximately 750,000 characters of protocol-relevant content spanning 180 distinct commands.

For integrating this knowledge, we use background-augmented prompting. The retrieved knowledge \( c \) is combined with task instructions \( q \) (e.g., “Generate a Modbus/TCP request for Function Code 03”) to form a comprehensive prompt, enabling protocol-compliant test case generation without fine-tuning the LLM. This method is computationally efficient and compatible with closed-source models.

\subsubsection{Quantized Low-Rank Adaptation (QLoRA)}
QLoRA \cite{ref37} enables memory-efficient fine-tuning of pre-trained LLMs by introducing low-rank decomposition layers, reducing computational and storage overhead. This is essential for adapting large models to industrial fuzz testing tasks in MALF.
We quantize the base model (Llama3.1-8B) to 4-bit precision, reducing memory usage by 75\% while maintaining performance \cite{ref38}. This allows efficient deployment on industrial hardware with minimal overhead.
QLoRA introduces trainable low-rank matrices into Transformer layers to adapt the model for fuzz testing. For example, the Seed Generation Agent retrieves protocol-specific structures, while the Test Case Generation Agent generates mutations. The transformation of the adapted weight matrix $W^{\prime}$ is expressed as:

\begin{equation}
\label{adapted_weights}
W' = W + \Delta W, \quad \Delta W = W_A W_B,
\end{equation}

\begin{equation}
\label{optimization_objective}
\mathcal{L} = -\sum_{i} \log P(y_i | x_i; \Theta_{\text{base}}, \Delta W),
\end{equation}

where $W_A \in \mathbb{R}^{d \times r}, \quad W_B \in \mathbb{R}^{r \times d}, \quad r \ll d$, $\Theta_{\text{base}}$ represents the frozen parameters of the pre-trained model, $\Delta W$ denotes the learnable low-rank adjustment, $x_i$ represents the input, and $y_i$ represents the target output.

QLoRA brings significant advantages to protocol fuzz testing by combining memory efficiency with rapid task-specific adaptation. By quantizing weights to 4-bit precision, GPU memory usage is reduced by up to 75\%, enabling efficient deployment even on industrial-grade hardware. This approach allows fine-tuning of specific transformer layers with minimal task-specific data, while preserving the general knowledge of the base model, ensuring both computational efficiency and task adaptability. Within MALF, QLoRA empowers each agent with specialized capabilities: the Seed Generation Agent reconstructs protocol-compliant seeds by leveraging RAG-retrieved contexts, the Test Case Generation Agent produces diverse and protocol-valid mutations to cover structural and semantic anomalies, and the Feedback Analysis Agent dynamically interprets system responses to refine mutation strategies. This synergy of efficiency and specialization positions QLoRA as a cornerstone of MALF’s scalability and precision.

By combining QLoRA fine-tuning with RAG-augmented prompting, MALF achieves a fine balance between domain-specific adaptability and computational efficiency. This integration allows the framework to dynamically adjust to new protocols and evolving industrial scenarios, setting a new benchmark for intelligent fuzz testing solutions.

\subsection{Seed Generation Agent}
The Seed Generation Agent is essential in transforming raw protocol traffic into structured, protocol-compliant fuzzing seeds. It automates this process by leveraging real-time traffic capture, dynamic protocol knowledge retrieval, and fine-tuned LLMs with Chain-of-Thought (CoT) reasoning\cite{ref39}, enhancing the efficiency and accuracy of fuzz testing.

\subsubsection{Functionality and Purpose}
The agent automates the extraction and validation of protocol-specific data to generate high-quality fuzzing seeds. Through CoT reasoning and RAG, it performs the following tasks:

\romannumeral1.\ Identify protocol types and characteristics: Determine ICPs based on traffic features.

\romannumeral2.\ Extract and validate fields: Extract essential fields such as function codes, addresses, and payload lengths while ensuring compliance with protocol rules.

\romannumeral3.\ Generate structured seed data: Create protocol-compliant seed data for use in fuzz testing, ensuring high reliability and vulnerability detection potential.

This systematic approach ensures that generated seeds are reliable, compliant, and effective for identifying vulnerabilities.

\subsubsection{Workflow Description}

\begin{figure}[!h]
\centering
\includegraphics[width=3 in]{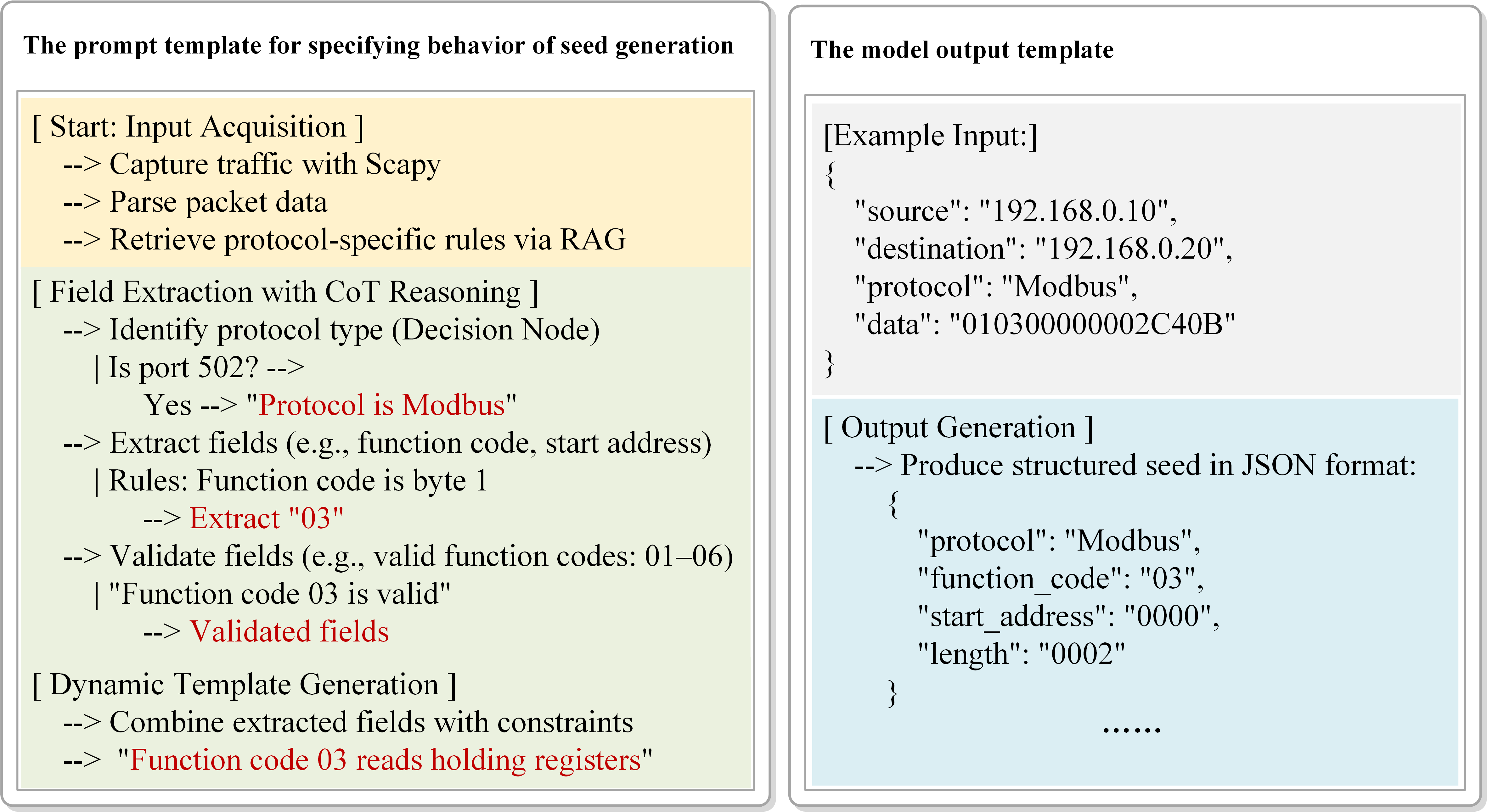}
\caption{The prompt template for specifying behavior of seed generation.}
\label{fig_3}
\end{figure}

The Seed Generation Agent follows a structured process to achieve its goals. First, traffic is captured using tools—Scapy, where raw payloads and metadata are parsed. Protocol-specific rules are retrieved from a knowledge base via RAG, providing necessary context for accurate parsing and validation. CoT reasoning then guides field extraction by breaking down tasks into logical steps, ensuring validation of fields (e.g., function code, address) against RAG-retrieved rules. Once fields are extracted, dynamic templates are created to encode protocol constraints, such as generating Modbus requests based on the extracted fields. Finally, the agent outputs the seed data in JSON format for fuzz testing. The prompt template of seed generation is shown in Fig. \ref{fig_3}.

\subsubsection{Integration of CoT and RAG}
CoT reasoning enables the decomposition of complex extraction tasks into manageable, sequential steps, ensuring logical consistency and minimizing errors. Each step in the CoT framework validates intermediate results by leveraging retrieved protocol-specific rules, thereby guiding subsequent actions. For example, during Modbus packet parsing, CoT ensures that fields such as function code and address are validated before proceeding to extract the length field. RAG enhances the system by dynamically retrieving relevant protocol knowledge, providing essential context for the LLM. This ensures that field extractions adhere to protocol-specific rules and constraints. RAG optimizations include efficient query construction targeting high-priority fields, hierarchical structuring of the knowledge base for rapid retrieval, and selective context injection to maintain prompt efficiency while maximizing relevance. Additionally, Template Automation is employed, where the agent automatically constructs protocol-specific templates by integrating RAG-retrieved rules with predefined logical patterns. These templates streamline the seed generation process, ensuring both accuracy and adaptability across diverse protocols.

\subsection{Test Case Generation Agent}
The Test Case Generation Agent converts high-quality seeds into diverse, protocol-compliant test cases by applying field, structural, and semantic mutations. Using dynamic prompt engineering, the agent uncovers vulnerabilities in the System Under Test (SUT) efficiently.

\subsubsection{Functionality and Objectives}
The agent's primary goal is to generate a broad set of test cases that evaluate the robustness of the SUT by introducing controlled variations. It achieves this through:
field mutations, which involve altering protocol field values such as function codes and data lengths; structural mutations, which modify the protocol’s architecture by inserting illegal fields or omitting essential ones; and semantic mutations, which create inconsistencies within the protocol data, such as mismatched lengths and payloads.

Additionally, the agent adjusts its mutation strategy based on feedback from the Feedback Analysis Agent, ensuring the test cases evolve in response to real-time insights about the SUT's vulnerabilities and behavior.

\begin{figure}[!h]
\centering
\includegraphics[width=3 in]{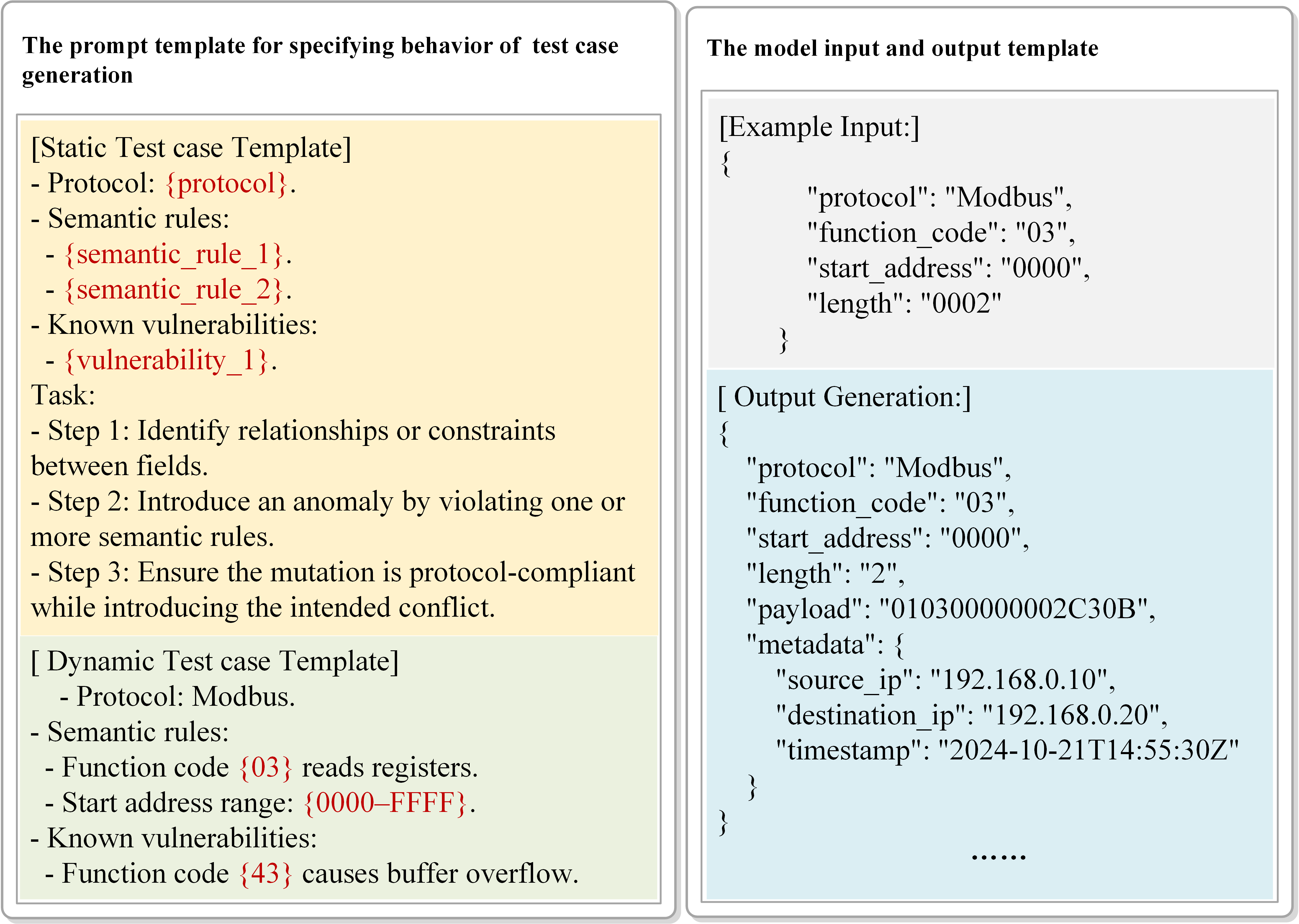}
\caption{The prompt template for specifying behavior of test case generation.}
\label{fig_4}
\end{figure}

\subsubsection{Workflow Description}

The test case generation workflow begins with Input Acquisition, where the agent receives structured seed data from the Seed Generation Agent and mutation priorities from the Feedback Analysis Agent. This provides the necessary context for generating test cases that target protocol vulnerabilities.

In the Mutation Logic phase, the agent applies a combination of field, structural, and semantic mutations. Field Mutations involve randomly modifying field values based on predefined rules, It can be formalized as: 
\begin{equation}
\label{field_mutation}
V' = V + \Delta V,
\end{equation}
where $V$ is the original field value and $\Delta V$ is a random or strategy-driven increment. 

Structural Mutations adjust the protocol’s structure by inserting illegal fields or removing mandatory ones, represented by $P^{\prime}$:
\begin{equation}
\label{structural_mutation}
P^{\prime} = P \cup F_{\text{insert}} \setminus F_{\text{delete}},
\end{equation}
where $P$ is the original packet, $F_{\text{insert}}$ is the set of inserted illegal fields, and $F_{\text{delete}}$ is the set of deleted mandatory fields. 

Semantic Mutations introduce inconsistencies within the protocol data, the mismatched field relationships, modeled by $S^{\prime}$:
\begin{equation}
\label{semantic_mutation}
S^{\prime} = f(S, \Delta S),
\end{equation}
where $S$ represents the original semantic relationships, and $\Delta S$ denotes the introduced anomalies.

Dynamic Strategy Adjustment dynamically modifies the mutation density based on feedback, following the equation:
\begin{equation}
\label{dynamic_strategy}
\rho = \rho_0 \cdot (1 + \alpha \cdot \text{feedback\_score}),
\end{equation}
where $\rho_0$ is the initial mutation density, $\alpha$ is the feedback influence weight, and $\text{feedback\_score}$ indicates the severity of detected anomalies. 

Dynamic prompts play a key role in guiding the LLM to perform precise mutations. The prompt template of test case generation is shown in Fig. \ref{fig_4}. It ensures semantic mutations are contextually aware, introducing meaningful anomalies. The core principles of dynamic prompt engineering are:

\romannumeral1. \ Task clarity: Clearly defining the mutation’s intent (e.g., mismatched fields).

\romannumeral2. \ Context injection: Using RAG to retrieve relevant protocol rules.

\romannumeral3. \ Logical decomposition: Breaking down mutation tasks into structured steps using CoT reasoning.

\romannumeral4. \ Flexibility: Adapting prompts dynamically for different mutation types.

In the Output Generation stage, the mutated test cases are formatted in JSON and hexadecimal format, making them ready for injection into the SUT.

\begin{algorithm}[h]
\caption{Prompt Engineering of Test Case Mutation Agent}
\label{alg:mutation_agent}
\begin{algorithmic}
\STATE \textbf{Input:} KnowledgeBase $K$
\STATE \textbf{Output:} $mutated\_seed$

\STATE

\STATE \textsc{Function} \textbf{CAPTURE\_SEED}()
\STATE \hspace{0.5cm} $seed \gets \{ \text{protocol, function\_code,start\_address, length} \}$
\STATE \hspace{0.5cm} \textbf{return} $seed$
\STATE
\STATE \textsc{Function} \textbf{RETRIEVE\_CONTEXT}($protocol$)
\STATE \hspace{0.5cm} $query \gets \text{"Protocol rules"} + protocol$
\STATE \hspace{0.5cm} $context \gets K.\textsc{SEARCH}(query, k=1)$
\STATE \hspace{0.5cm} \textbf{return} $context$
\STATE
\STATE \textsc{Function} \textbf{CONSTRUCT\_PROMPT}($seed, context$)
\STATE \hspace{0.5cm} $prompt \gets \text{"Context:n}" + context + \text{"n"}$
\STATE \hspace{0.5cm} $prompt \gets prompt + \text{"Task:}"$ 
\STATE \hspace{1.2cm} $\text{"- Step 1: Perform field mutations.}"$ 
\STATE \hspace{1.2cm} $\text{"- Step 2: Perform structural mutations.}"$ 
\STATE \hspace{1.2cm} $\text{"- Step 3: Perform semantic mutations.}"$ 
\STATE \hspace{0.5cm} $\text{"Input seed:}" + seed$
\STATE \hspace{0.5cm} \textbf{return} $prompt$
\STATE

\STATE \textsc{Function} \textbf{EXECUTE\_PROMPT}($prompt$)
\STATE \hspace{0.5cm} $M \gets \textsc{LLM}(\text{"fine-tuned-LLM"})$
\STATE \hspace{0.5cm} $output \gets M.\textsc{GENERATE}(prompt)$
\STATE \hspace{0.5cm} \textbf{return} $output$
\STATE

\STATE $seed \gets \textsc{CAPTURE\_SEED}()$
\STATE $context \gets \textsc{RETRIEVE\_CONTEXT}(seed.\text{protocol})$
\STATE $prompt \gets \textsc{CONSTRUCT\_PROMPT}(seed, context)$
\STATE $mutated\_seed \gets \textsc{EXECUTE\_PROMPT}(prompt)$

\STATE \textbf{return} $mutated\_seed$
\end{algorithmic}
\label{alg1}
\end{algorithm}

\subsection{Feedback Analysis Agent}
The Feedback Analysis Agent analyzes responses from the SUT and dynamically adjusts the mutation strategies of the Test Case Generation Agent. This enables a seamless, automated fuzz testing loop driven by real-time feedback and adaptive strategy refinement.

\subsubsection{Functionality and Objectives}
The agent’s primary tasks are:

\romannumeral1. \ Real-Time Feedback Collection: Captures SUT responses, categorizing them into normal, abnormal, and critical anomalies (e.g., crashes, timeouts).

\romannumeral2. \ Anomaly Analysis: Uses weighted scoring to assess anomaly severity and extract potential vulnerability patterns.

\romannumeral3. \ Dynamic Strategy Adjustment: Refines mutation density, prioritizes high-risk fields, and adjusts mutation directions based on the feedback, optimizing test coverage and vulnerability detection.

\subsubsection{Workflow Description}
The process begins with the collection of SUT feedback, including timeouts, error codes, and normal acknowledgments, as well as historical test case data. This allows for the identification of trends and patterns that may not be immediately visible from individual results.
The collected feedback is then classified into three categories: Normal Responses, which include protocol-compliant acknowledgments; Abnormal Responses, such as unexpected error codes or delays; and Critical Anomalies, which encompass severe issues like timeouts, crashes, or resource exhaustion. This classification enables the prioritization of anomalies based on their potential impact on the agent.

To quantify the severity of detected anomalies, a weighted scoring system is employed: \begin{equation}
\label{severity_score}
S = w_1 \cdot E + w_2 \cdot T + w_3 \cdot R,
\end{equation}
where $E$ represents the anomaly type score (e.g., crash = 10, timeout = 8), $T$ denotes the response time score (e.g., timeout = 8, delay = 3), and $R$ signifies the resource utilization score (e.g., CPU/memory spikes). The weights $w_1$, $w_2$, and $w_3$ are adjusted according to the testing objectives.

The strategy is dynamically adjusted based on feedback severity. The mutation density, $\rho^{\prime}$ is updated using the formula:
\begin{equation}
\label{mutation_density}
\rho^{\prime} = \rho_0 \cdot \left( 1 + \beta \cdot \frac{S}{S_{\text{max}}} \right),
\end{equation}
where $\rho_0$ is the initial mutation density, $\beta$ is a scaling factor, and $\frac{S}{S_{\text{max}}}$ represents the normalized severity score. High-risk fields are prioritized, and mutation directions are adjusted to target areas with higher vulnerability, ensuring efficient and comprehensive testing.

\subsubsection{Prompt Engineering for Feedback Analysis and Strategy Adjustment}
To automate the translation of feedback into actionable mutation strategies, the Feedback Analysis Agent uses dynamically constructed prompts to guide the LLM in three core functions:

\romannumeral1. \ Classifying Feedback: The LLM categorizes feedback (normal, abnormal, or critical anomalies) based on predefined criteria and historical data patterns.

\romannumeral2. \ Assigning Severity Scores: The LLM assigns severity scores using a weighted system that accounts for anomaly type, response time, and resource usage, prioritizing issues with the highest potential impact.

\romannumeral3. \ Proposing Strategy Adjustments: Based on feedback and severity scores, the LLM adjusts mutation strategies by modifying mutation density, reordering focus to high-risk fields, and optimizing directions to target vulnerable patterns.

\begin{figure}[!h]
\centering
\includegraphics[width=3 in]{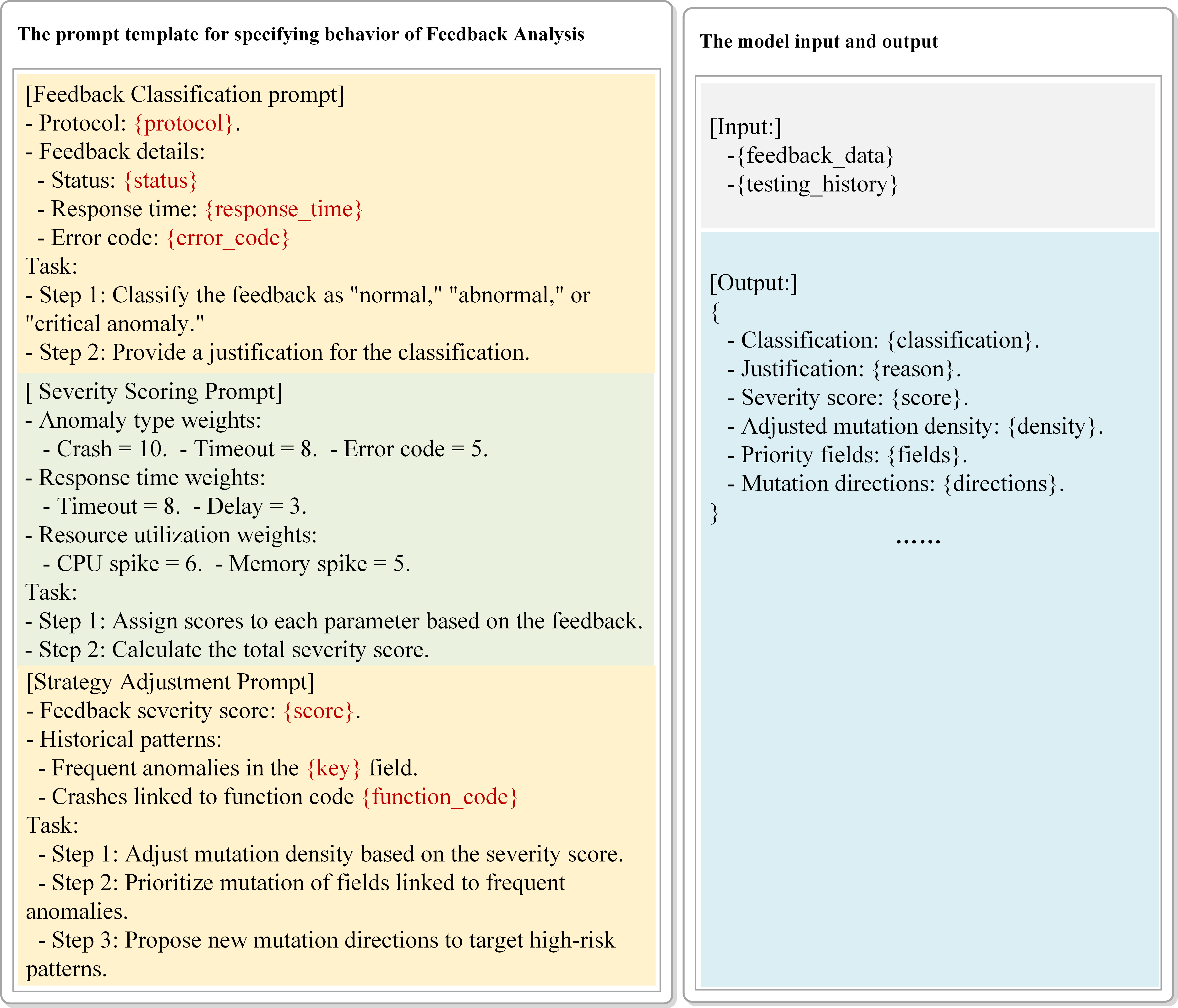}
\caption{The prompt template for specifying behavior of Feedback Analysis Agent.}
\label{fig_5}
\end{figure}

The prompt template used by the Feedback Analysis Agent (shown in Fig. \ref{fig_5}) integrates knowledge bases and feedback loops to refine strategies. This starts with Knowledge Base Retrieval, where historical anomalies and strategies inform current adjustments, ensuring lessons learned improve future testing.
For evaluating complex feedback, Fuzzy Logic Optimization is employed. Fuzzy rules assess conditions like "IF anomaly frequency is high AND stability is low, THEN reduce mutation density," enabling the system to adapt dynamically to a range of scenarios.
The process concludes with Iterative Improvement, where the knowledge base is continually updated with new insights and strategies. As the system encounters novel anomalies, it incorporates this information, evolving its approach and improving vulnerability detection over time. This creates a closed-loop learning system that refines the agent’s effectiveness in addressing critical vulnerabilities.

\subsection{Communication Interaction Module}
The Communication Interaction Module ensures seamless real-time data exchange, test case injection, and inter-agent communication within MALF. It addresses challenges in parallelization, resilience, and scalability, improving system reliability and enabling high-throughput testing across diverse protocols.

\subsubsection{Module Functionality}
This module performs several critical functions:
Real-Time Traffic Capture: It monitors protocol-specific traffic between the SUT and external devices, parsing application-layer data for test case generation.

\romannumeral1. \ Test Case Injection: Mutated test cases are injected into communication flows, with responses captured for feedback analysis.

\romannumeral2. \ Inter-Agent Coordination: Task queues and asynchronous messaging synchronize agent workflows, ensuring data consistency and operational efficiency.

\romannumeral3. \ Fault-Tolerant Operation: Error handling, retry mechanisms, and fallback strategies maintain reliability during adverse conditions.

\romannumeral4. \ Extensible Design: A modular architecture allows easy integration of new protocols, fuzzing techniques, and reporting modules for future scalability.

\subsubsection{Key Considerations in Multi-Agent Coordination Mechanisms}

Efficient multi-agent coordination is essential for optimal fuzz testing performance. Parallelization boosts throughput by running multiple tasks—such as test case generation and injection—concurrently. Load balancing dynamically distributes resources, preventing bottlenecks, while module-specific parallelism allows simultaneous protocol knowledge retrieval and test case generation. This enables faster, more efficient testing by injecting multiple test cases into various SUT instances at once.

Equally important is resilience. To ensure continuous operation, even under failure conditions, the system employs robust error handling and retry mechanisms, such as exponential backoff, to recover from failures. Health monitoring tracks module status with heartbeat signals and centralized logging, allowing the system to gracefully degrade when components fail, ensuring critical operations continue without disruption. For instance, if knowledge retrieval fails, generic mutation strategies are used to maintain progress.

Finally, extensibility is key for adapting to new protocols or fuzzing techniques. A modular architecture supports the seamless integration of new components without disrupting the system. The plugin system allows for dynamic addition of protocol parsers or fuzzing methods, and configurable modules offer granular control through configuration files. This ensures the system can easily scale and evolve with minimal disruption, such as when adding support for a new protocol through a specialized parsing module.

\begin{algorithm}[h]
\caption{Multi-Agent Coordination with ZeroMQ}\label{alg:multi_agent_coordination}
\begin{algorithmic}
\STATE \textbf{Data Structures:}
\STATE \hspace{0.5cm} $context$: A ZeroMQ Context for message passing
\STATE \hspace{0.5cm} $seeds\_queue, test\_cases\_queue, responses\_queue$: ZeroMQ sockets (PUB/SUB)

\STATE \textbf{Initialization:}
\STATE \hspace{0.5cm} $context \gets \text{new Context()}$
\STATE \hspace{0.5cm} $seeds\_queue \gets context.\text{socket(PUB)}$
\STATE \hspace{0.5cm} $test\_cases\_queue \gets context.\text{socket(PUB)}$
\STATE \hspace{0.5cm} $responses\_queue \gets context.\text{socket(PUB)}$

\STATE \textbf{Function} \textsc{SEED\_GENERATION}()
\STATE \hspace{0.5cm} $seeds\_queue.\text{bind}("tcp://*:5555")$
\STATE \hspace{0.5cm} \textbf{while True do}
\STATE \hspace{1.0cm} $seed\_data \gets \text{CAPTURE\_TRAFFIC()}$ 
\STATE \hspace{1.0cm} $seeds\_queue.\text{send\_json}(\{ \text{"event": "seed","data"} \})$

\STATE \textbf{Function} \textsc{TEST\_CASE\_GENERATION}()
\STATE \hspace{0.5cm} $test\_cases\_queue.\text{bind}("tcp://*:5556")$
\STATE \hspace{0.5cm} $seeds\_queue.\text{connect}("tcp://localhost:5555")$
\STATE \hspace{0.5cm} \textbf{while True do}
\STATE \hspace{1.0cm} $seed\_msg \gets seeds\_queue.\text{recv\_json}()$
\STATE \hspace{1.0cm} $seed \gets seed\_msg.\text{data}$
\STATE \hspace{1.0cm} $test\_case \gets \text{GENERATE\_TEST\_CASE}(seed)$ 
\STATE \hspace{1.0cm} $test\_cases\_queue.\text{send\_json}(\{ \text{"event": "test\_case",}$
\STATE \hspace{4.0cm} \text{"data": test\_case} \})

\STATE \textbf{Function} \textsc{FEEDBACK\_ANALYSIS}()
\STATE \hspace{0.5cm} $responses\_queue.\text{connect}("tcp://localhost:5557")$
\STATE \hspace{0.5cm} \textbf{while True do}
\STATE \hspace{1.0cm} $response\_msg \gets responses\_queue.\text{recv\_json}()$
\STATE \hspace{1.0cm} $\text{UPDATE\_STRATEGIES}(response\_msg)$

\STATE \textbf{Function} \textsc{HEARTBEAT}($agent\_id$)
\STATE \hspace{0.5cm} \textbf{while True do}
\STATE \hspace{1.0cm} $\text{SEND\_HEARTBEAT}(agent\_id)$ 
\STATE \hspace{1.0cm} $\text{SLEEP}(5)$

\STATE \textbf{Function} \textsc{REDISTRIBUTE\_TASKS}($failed\_agent$)
\STATE \hspace{0.5cm} \text{// reassign tasks from $failed\_agent$ to active agents}
\STATE \hspace{0.5cm} \text{// update routing tables or load-balance existing queues}

\end{algorithmic}
\end{algorithm}

\subsubsection{Enhanced Implementation with Coordination Mechanisms}

To implement efficient multi-agent coordination, a system is designed where each agent operates independently as a process or thread, communicating via ZeroMQ PUB/SUB sockets, as shown in Algorithm 2.

The Seed Generation Agent publishes captured traffic data to a dedicated port (e.g., tcp://:5555), allowing the Test Case Generation Agent to subscribe and receive the data in real-time. Upon receiving the seed, the Test Case Generation Agent mutates the data and publishes the resulting test cases on another port (e.g., tcp://:5556). Downstream agents, like the Feedback Analysis Agent, subscribe to this feed, processing the test cases asynchronously. Simultaneously, the Feedback Analysis Agent listens to system feedback (e.g., SUT output) via a separate port (tcp://*:5557) to refine strategies or request new seeds.
To ensure reliability, each agent periodically sends heartbeat signals (e.g., "SeedAgent alive") to a central monitoring service. If a heartbeat is missed, the system triggers a REDISTRIBUTE\_TASKS function to reassign tasks to active agents. Since PUB/SUB sockets naturally support one-to-many communication, reassignment involves rerouting data without altering the underlying logic.
This modular architecture supports seamless scalability. New protocol handlers can be added as independent agents with their own PUB/SUB pairs, integrating into the existing system through JSON messages. This design isolates each process, reducing the risk of failure and enabling flexible, cross-language development on different hosts.

\section{Evaluation}
To evaluate the effectiveness of MALF, we seek to answer the following key questions:
\begin{itemize}
    \item {Q1: \textit{How effectively does the MALF multi-agent coordination mechanism generate high-quality seeds and diverse test cases targeting ICP vulnerabilities?} }
    \item {Q2: \textit{How do domain augmentation and efficient fine-tuning mechanisms amplify fuzzing performance?}}
    \item {Q3: \textit{How does the framework perform in real-world ICSs, and what vulnerabilities can it uncover that traditional approaches miss?}}
\end{itemize}

\subsection{Experimental Setup}
\subsubsection{Computational Environment}
The model fine-tuning phase is conducted on a high-performance cluster featuring 8 NVIDIA H100 GPUs, an Intel Xeon Gold 4310 CPU (12C24T, 2.1 GHz), and 128 GB DDR4 memory, supporting large-scale training and inference. The testing and deployment phase utilizes 2 NVIDIA 4090 GPUs, an Intel Core i9-14900K CPU (24C32T, 2.4 GHz), and 128 GB DDR5 memory, optimized for high-throughput inference and parallel task execution. The framework is built on Python 3.9+, using Hugging Face Transformers for LLM functions, Scapy for network traffic manipulation, and Redis for state management.

\subsubsection{Fuzz Testing Platform}
\begin{figure}[!h]
\centering
\includegraphics[width=3 in]{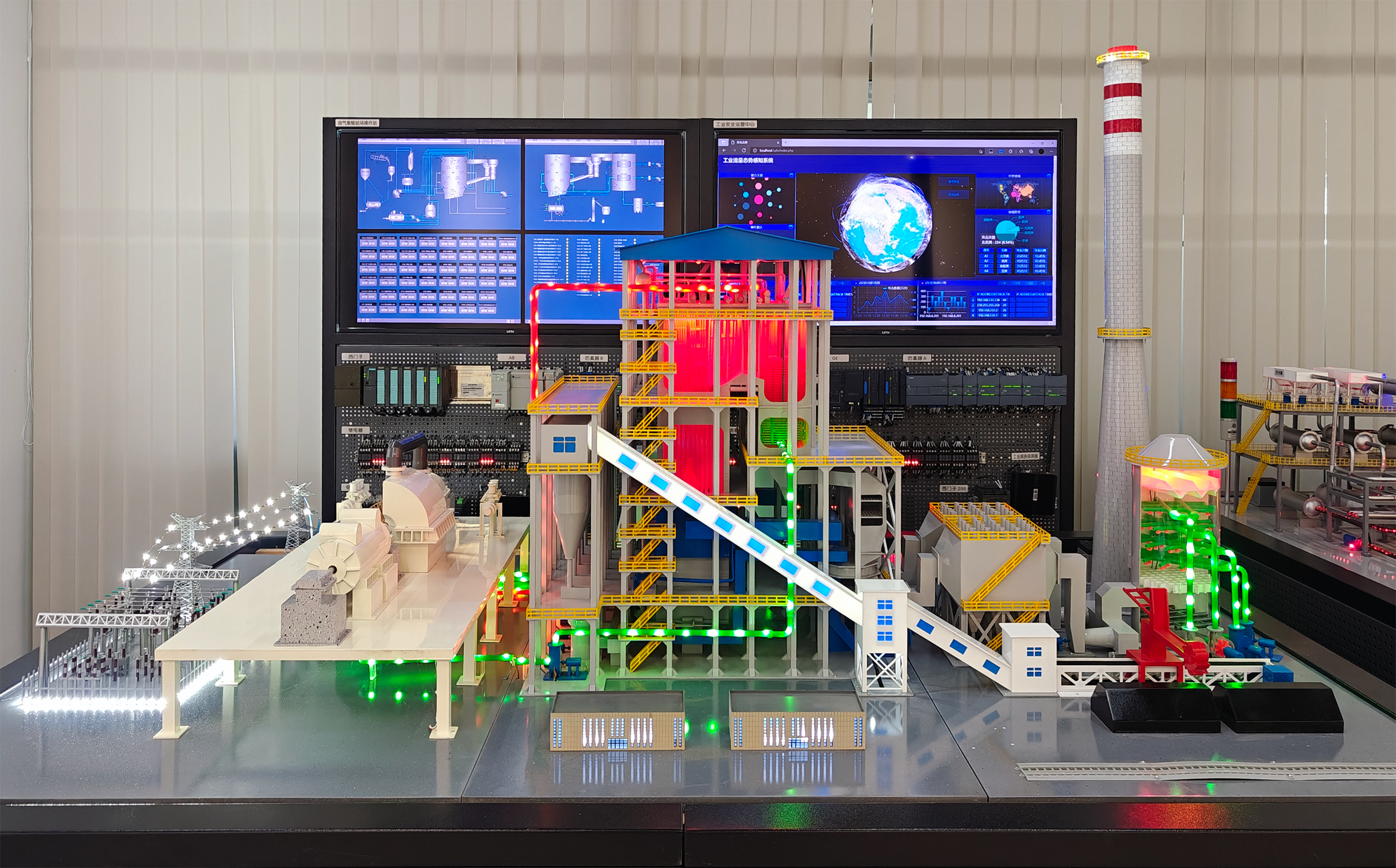}
\caption{Experimental platform: Industrial Attack-Defense Range for power plant.}
\label{fig_6}
\end{figure}

\begin{figure}[!t]
\centering
\includegraphics[width=3 in]{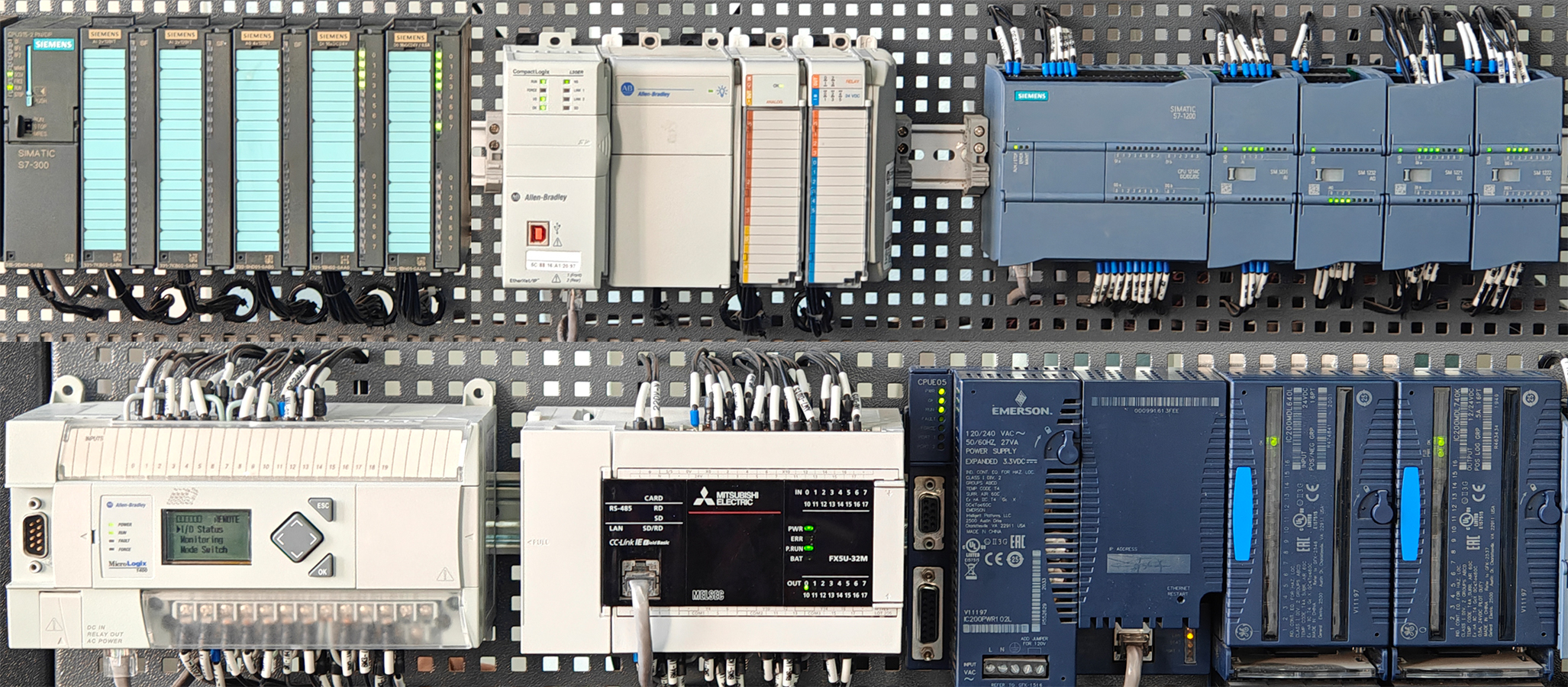}
\caption{System under test equipment.}
\label{fig_7}
\end{figure}

Experiments are conducted in the Industrial Attack-Defense Range for power plant security, located within the Key Laboratory of Information Security in the Petroleum and Chemical Industry of Liaoning Province. As shown in Fig. \ref{fig_6} -\ref{fig_7}, the range integrates the current mainstream commercial PLC and provides a comprehensive test platform for real-world ICS.

Unlike traditional fuzzing tools and benchmarks such as FuzzBench or OSS-Fuzz, which fall short in ICS scenarios, this hybrid fuzzing platform combines both simulated and real hardware, providing a more accurate assessment of protocol vulnerabilities and improving fuzzing performance validation. The specific hardware details are shown in Table \ref{tab:table1}. 

\subsubsection{Baseline Methods}
To evaluate our Multi-Agent LLM Fuzzing Framework, we compare it against three established fuzzing solutions representing varying levels of protocol awareness and automation:

\textbf{Peach} \cite{ref17}: A widely used fuzzing framework focused on input data mutation. While it supports basic protocol models, Peach lacks deep industrial protocol-specific logic. We followed Peach’s guidelines to create models for each target service, defining data formats and state transitions.

\textbf{ChatAFL} \cite{ref29}: An LLM-based fuzzing tool using a single-agent architecture. Unlike our multi-agent framework, it lacks dynamic coordination and context retrieval. We adapted ChatAFL by utilizing its built-in prompting mechanism and seed data derived from ICS traffic logs.

\textbf{NCMFuzzer} \cite{ref40}: A specialized tool for ICS protocols, designed to handle packet structures like Modbus/TCP and S7. It integrates domain-specific knowledge to uncover protocol vulnerabilities. We provided captured network traces and relevant field constraints (e.g., function codes, addresses) for configuration.

Each baseline fuzzer was configured per the recommended practices to ensure fairness. Generation-based fuzzers (Peach) required hand-crafted models, while mutation-based (NCMFuzzer) and LLM-based (ChatAFL) fuzzers were initialized with seed inputs from ICS environments. This setup allowed a fair evaluation of how our multi-agent approach improves over conventional or single-agent fuzzers in detecting complex ICS vulnerabilities.

\begin{table}[H]
\caption{Details of the device under test\label{tab:table1}}
\centering
\renewcommand\arraystretch{1.3}
\begin{tabularx}{\textwidth}{@{} XXXX@{} }
\toprule
Manufacturer & Monitor & Device & Protocol \\ 
\midrule
SIEMENS & Step7  & (1) S7-300   & S7comm \\ 
SIEMENS & Portal V13  & (2) S7-1200 & S7comm \\ 
Rockwell & RSLogix 5000 & (3)1766-L32BWA & Modbus/TCP \\ 
Mitsubishi & GX works3 & (4)FX5U-32MT & Modbus/TCP \\ 
Rockwell & RSLogix 5000 & (5)1769-L30ER & Ethernet/IP \\ 
Emerson & Proficy machine & (6)VersaMax IC200 & Ethernet/IP \\ 
\bottomrule
\end{tabularx}
\end{table}

\subsubsection{Parameter Settings}
The Multi-Agent LLM Fuzzing Framework is optimized for efficiency and adaptability. It uses Llama3.1-8B fine-tuned via LoRA with 4-bit quantization, ensuring low memory usage and high performance.
The model is trained for 3–5 epochs, using a batch size of 16 and a learning rate of $1 \times 10^{-4}$, with a warm-up phase to avoid overfitting. During inference, the temperature is set to 0.7, with top-k = 50 and top-p = 0.95 to balance exploration with protocol compliance.

Each agent is fine-tuned for specific tasks: the Seed Generation Agent retrieves protocol knowledge from a RAG database (including protocol specs, CVEs, and traces), ensuring accurate seed generation. The Test Case Generation Agent uses CoT reasoning for mutation creation, adjusting temperature for deterministic or diverse mutation logic. The Feedback Analysis Agent processes SUT responses, assigns severity scores, and refines future mutation strategies.
The RAG database integrates protocol documentation and ICS vulnerability data, optimizing context retrieval with a 0.85 similarity threshold and a 512-token context size, ensuring accurate and efficient agent workflows.

\subsection{Evaluation Criteria}
This section evaluates MALF across two key dimensions: fuzzing effectiveness and test case richness. Fuzzing effectiveness assesses the framework's ability to generate impactful test cases that trigger meaningful system responses, while test case richness measures the diversity and coverage of generated test cases. Metrics within these dimensions provide quantitative insights into the framework’s performance, supported by visualizations that highlight trends under various workloads\cite{ref41,ref42}.

\subsubsection*{Fuzzing Effectiveness}

The \textbf{Test Case Pass Rate (TCPR)} quantifies the proportion of test cases successfully processed by the SUT without rejection or parsing errors. It is defined as:
\begin{equation}
\label{tcpr_definition}
\text{TCPR} = \frac{\#\bigl(\text{passed}\bigr)}{\#\bigl(\text{total}\bigr)},
\end{equation}
where $\#\bigl(\text{passed}\bigr)$ is the number of test cases accepted by the SUT, and $\#\bigl(\text{total}\bigr)$ is the total number of test cases generated (through mutation or generation).
A higher TCPR indicates that test cases adhere to the protocol's syntactic rules, facilitating meaningful system responses.

The \textbf{Exception Triggers Number (ETN)} measures the framework's ability to uncover vulnerabilities like crashes or unexpected states. In our setup, three fuzzing cycles (each lasting 24 hours) are used, and ETN is defined as the total number of crashes per cycle:
\begin{equation}
\label{etn_definition}
\mathrm{ETN}_i = \#\bigl(\text{crashes in cycle } i\bigr), \quad i = 1, 2, 3,
\end{equation}
$\mathrm{ETN}_i$ reflects the system's susceptibility to faults and its robustness in handling anomalies.

\subsubsection*{Test Case Richness}

We assess Seed Quality based on \textbf{Coverage}, which measures how many valid field-value combinations from the protocol specification are included in the seed corpus:
\begin{equation}
\label{coverage_definition}
\text{Coverage} = \frac{\#\bigl(\text{covered combos}\bigr)}{\#\bigl(\text{total combos}\bigr)},
\end{equation}
where $\#\bigl(\text{covered combos}\bigr)$ is the number of distinct field-value pairs present in the seeds, and $\#\bigl(\text{total combos}\bigr)$ is the total possible pairs defined by the protocol’s specification.
Higher coverage indicates a more comprehensive seed set, improving the chances of triggering diverse system behaviors.

To quantify the variability of field values across all generated test cases, we use the \textbf{Shannon Entropy}\cite{ref43} :
\begin{equation}
\label{shannon_entropy}
H(X) = -\sum_{i} p(x_i) \log_{2}\bigl(p(x_i)\bigr),
\end{equation}
where $X$ is a random variable representing field values encountered in the mutated test set, and $p(x_i)$ is the probability of observing a specific value $x_i$. A larger $H(X)$ indicates a wider distribution of field values, implying that mutations explore a more diverse range of possibilities. 
A larger $H(X)$ indicates a more diverse distribution of field values, suggesting that the mutations explore a broader range of possibilities.

\subsection{Results Analysis}
We now consolidate quantitative findings and interpretive insights across key metrics, illustrating how MALF outperforms baseline approaches, demonstrates domain-awareness, and pushes fuzz testing beyond traditional limits. Evaluation results are summarized in Tables \ref{tab:table2} and \ref{tab:table3}.

\subsubsection*{Fuzzing Effectiveness}
\begin{figure}[!h]
\centering
\includegraphics[width=3 in]{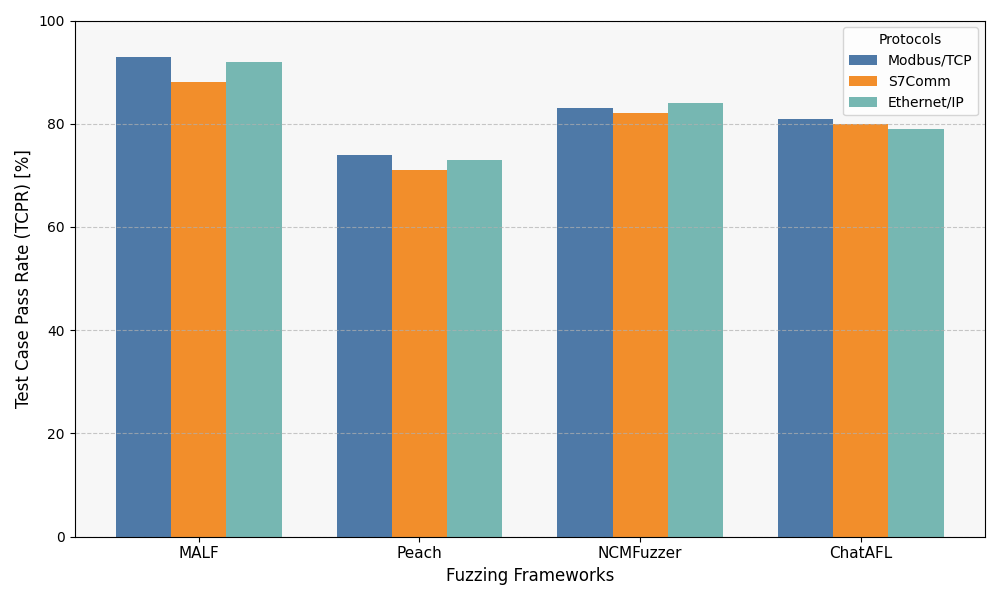}
\caption{Test Case Pass Rate (TCPR) Across Protocols and Frameworks.}
\label{fig_8}
\end{figure}

\begin{table*}[!t]
\caption{Evaluation results of TCPR and ETN}
\centering
\renewcommand{\arraystretch}{1.3}
\begin{tabular}{c|ccc|ccc}
\hline
\multirow{2}{*}{\textbf{Fuzzing Framework}} & \multicolumn{3}{c|}{\textbf{Test Case Pass Rate (TCPR), \%}} & \multicolumn{3}{c}{\textbf{Exception Triggers Number (ETN)}}  \\  \cline{2-7}
                           & \textbf{Modbus/TCP} & \textbf{S7Comm} & \textbf{Ethernet/IP} & \textbf{Modbus/TCP} & \textbf{S7Comm} & \textbf{Ethernet/IP} \\ \hline
\textbf{MALF}              & \textbf{92.02}     & \textbf{88.94}  & \textbf{91.15}       & \textbf{22}         & \textbf{17}     & \textbf{16}          \\
\textbf{NCMFuzzer}         & 83.09              & 80.92           & 84.50               & 17                  & 8               & 11                   \\
\textbf{ChatAFL}           & 82.15              & 81.85           & 80.75               & 13                  & 11              & 12                   \\
\textbf{Peach}             & 74.86              & 71.85           & 73.28               & 9                   & 5               & 8                    \\ \hline
\end{tabular}
\label{tab:table2}
\end{table*}

As shown in Fig. \ref{fig_8}. MALF achieves a remarkable TCPR of 88-92\% across protocols like Modbus/TCP, S7 and Ethernet/IP, significantly outperforming baselines such as Peach (74\%), NCMFuzzer (83\%), and ChatAFL (81\%). This high TCPR reflects MALF's robust multi-agent synergy, where the Seed Generation Agent ensures protocol-compliant seeds and the Test Case Generation Agent employs structured Chain-of-Thought CoT reasoning to maintain syntactic validity. By integrating RAG, MALF injects domain-specific knowledge into the fuzzing process, including protocol constraints and vulnerability disclosures. This targeted approach minimizes rejections and parsing failures, producing fewer wasted test cases and enhancing the framework's ability to probe industrial systems effectively.

\begin{figure}[!h]
\centering
\includegraphics[width=3in]{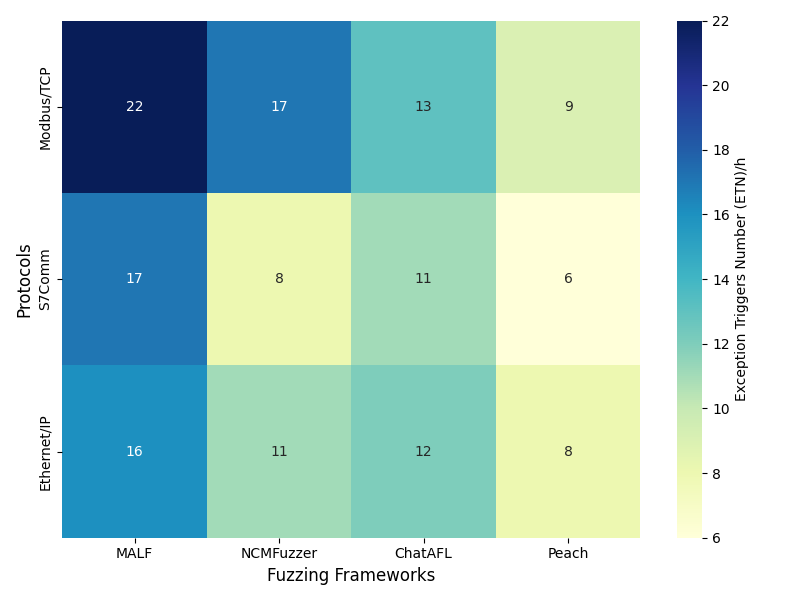}
\caption{Exception Triggers Number (ETN) Across Frameworks.}
\label{fig_9}
\end{figure}

In addition to its high TCPR, MALF demonstrates superior ETN as shown in Fig. \ref{fig_9} , across three 24-hour fuzzing cycles, MALF consistently induces around 22 crashes on a tested commercial PLC running Modbus/TCP—significantly outstripping NCMFuzzer (17), ChatAFL (15), and Peach (12). For S7 and Ethernet/IP , a similarly elevated ETN signals MALF’s knack for pinpointing critical system vulnerabilities. 

High Fuzzing Effectiveness are particularly telling in an industrial context: they imply that MALF’s iterative refinement is effective at eliciting corner-case anomalies and unanticipated failure states. By zeroing in on suspicious behaviors detected in prior runs, the agents exploit ICS intricacies—for instance, specialized command sequences or unusual parameter ranges—to systematically expose fragile code paths. Consequently, the robust synergy of domain knowledge, multi-agent coordination, and adaptive feedback cements MALF’s reputation as a formidable fuzzer for real-world industrial settings.

\subsubsection*{Test Case Richness}
\begin{figure*}[t]
\centering
\subfloat[]{\includegraphics[width=2in]{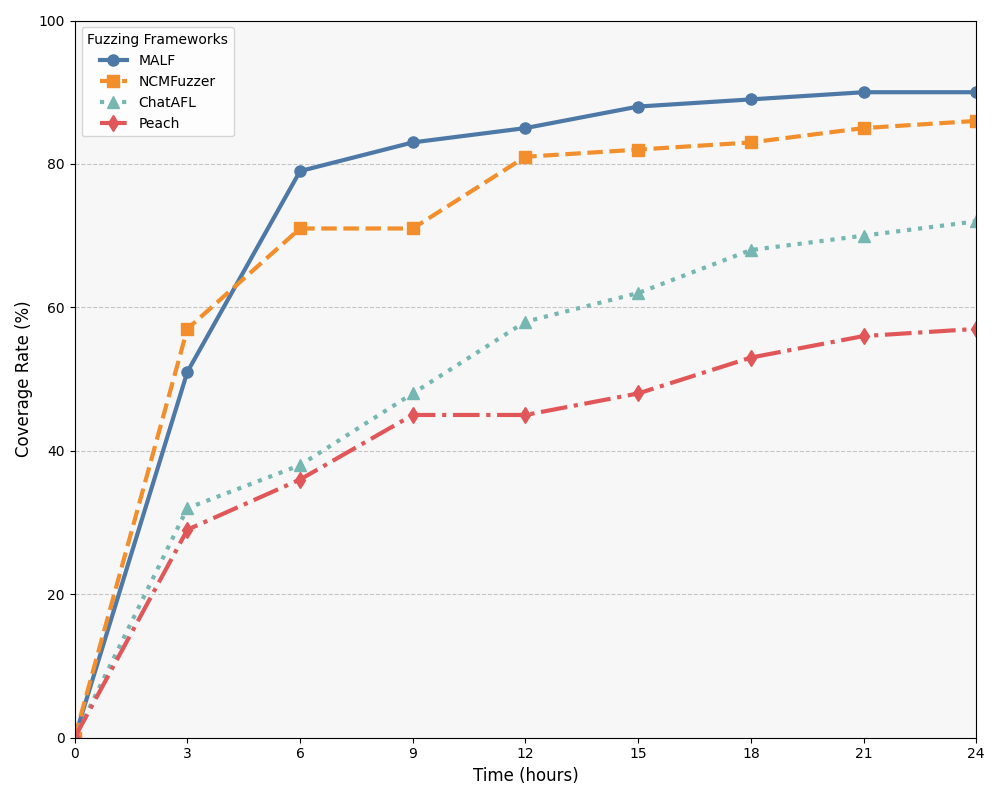}%
\label{fig_first_case}}
\hfil
\subfloat[]{\includegraphics[width=2in]{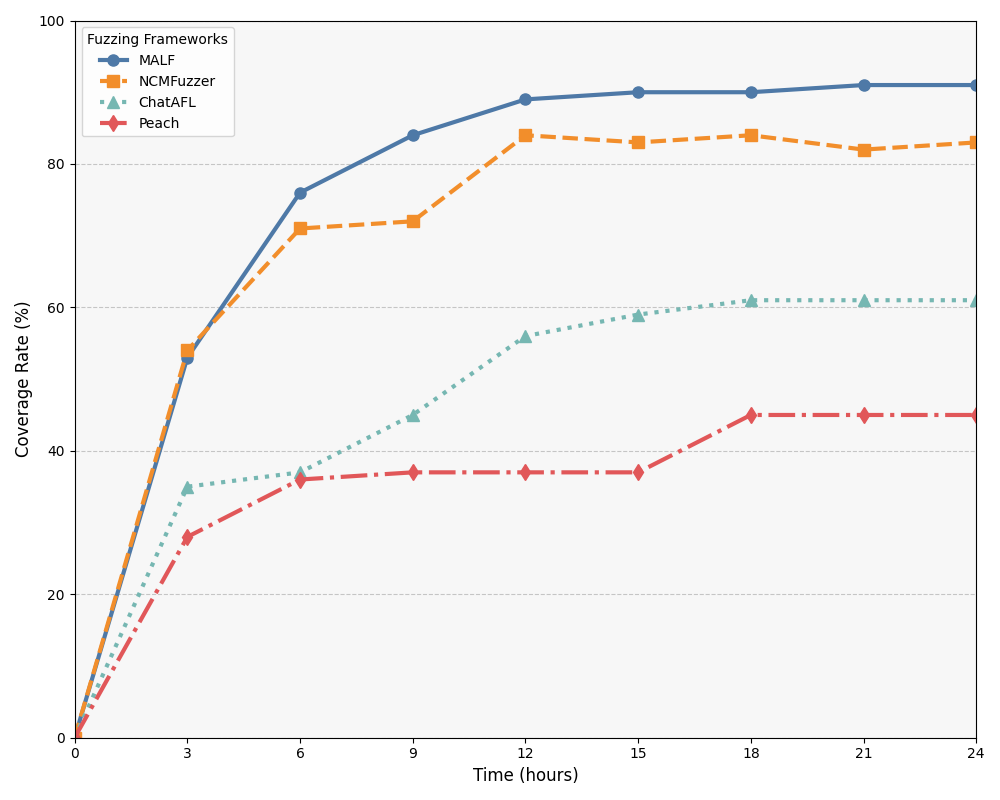}%
\label{fig_second_case}}
\hfil
\subfloat[]{\includegraphics[width=2in]{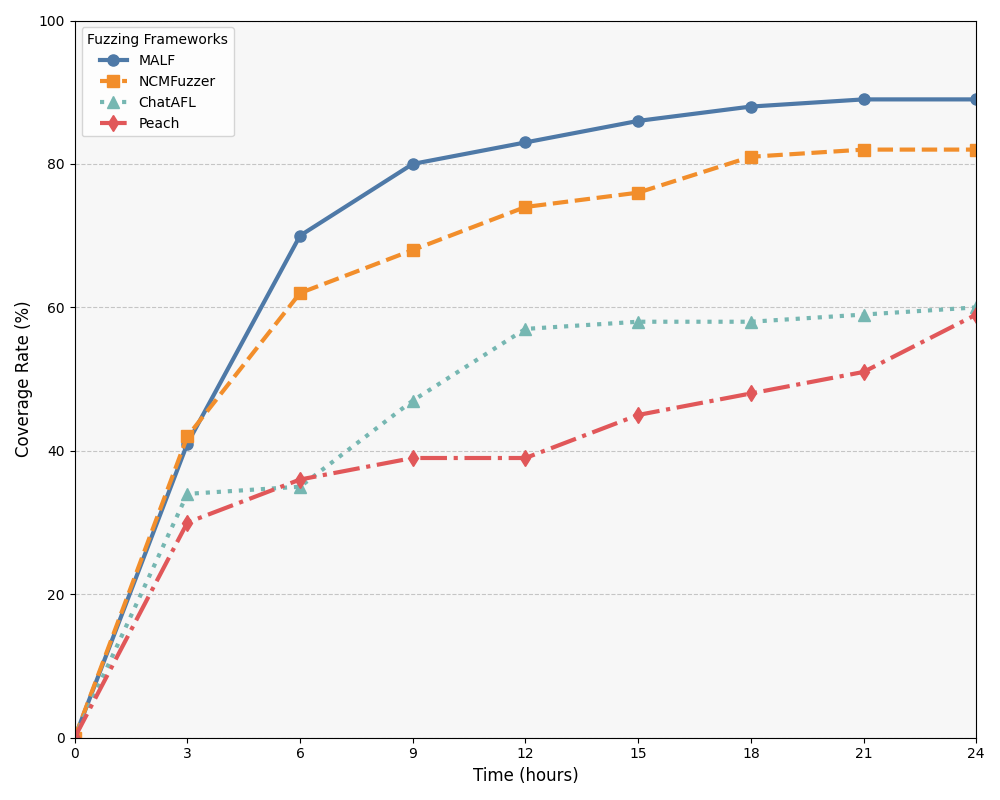}%
\label{fig_third_case}}
\caption{Coverage Rates Over Time aross Protocols and Frameworks (Sampled Every 3 Hours). (a) Modbus/TCP, (b) S7Comm, and (c) Ethernet/IP.}
\label{fig_10}
\end{figure*}

MALF excels in test case richness, achieving both comprehensive coverage and high mutation diversity, ensuring thorough ICP exploration.
As shown in Fig. \ref{fig_10}, MALF achieves over 90\% coverage, surpassing Peach, ChatAFL (60\%), and NCMFuzzer (83\%). This superior coverage reflects MALF's ability to systematically incorporate valid field-value combinations, including rarely used or vendor-specific protocol states. Powered by a versatile RAG database, the Seed Generation Agent retrieves nuanced protocol details, such as vendor constraints and ICS logs, ensuring that seeds represent diverse protocol functionalities. This capability enables MALF to probe seldom-invoked functionality, shining a spotlight on hidden vulnerabilities and latent design flaws that are often overlooked by less adaptive fuzzing tools.

\begin{figure}[!h]
\centering
\includegraphics[width=3in]{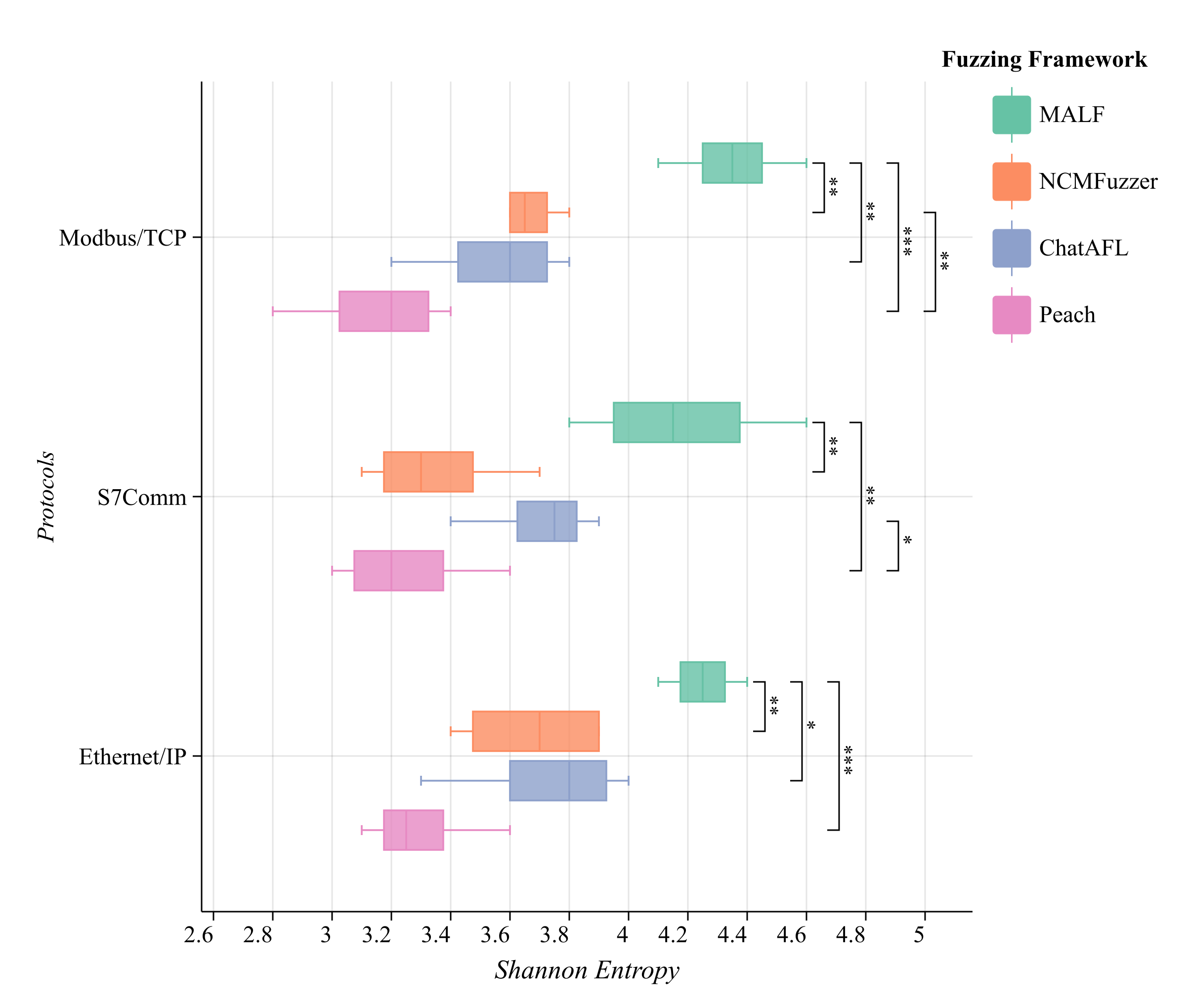}
\caption{Shannon Entropy Across Fuzzing Frameworks.}
\label{fig_11}
\end{figure}

In Fig. \ref{fig_11}, MALF also outperforms others in mutation diversity, with Shannon Entropy values of 4.2–4.6 bits, higher than Peach, NCMFuzzer, and ChatAFL (3.1–4.1 bits). This diversity stems from dynamic CoT prompting and real-time feedback adjustments, allowing the Test Case Generation Agent to explore a broad array of field values and structural permutations. By introducing semantic anomalies, varying mutation intensities, and adapting strategies based on SUT responses, MALF systematically expands the scope of protocol testing. This approach ensures that test cases venture into unexplored corners of protocol behavior, uncovering subtle yet high-impact vulnerabilities.

Through the experiments conducted above, we answer \textit{Q1} , the synergy between comprehensive coverage and mutation diversity amplifies MALF’s ability to test protocol-specific complexities. Seeds derived from genuine ICS logs and enriched by RAG retrieval maximize initial protocol representation, while dynamic mutations introduce variability necessary for uncovering elusive flaws. Together, these capabilities establish MALF as a powerful domain-oriented fuzzing solution that excels in generating diverse and impactful test cases, driving deeper exploration of industrial protocols and enhancing the likelihood of detecting critical vulnerabilities.

\begin{table*}[!t]
\caption{Evaluation results of Coverage and Shannon Entropy}
\centering
\renewcommand{\arraystretch}{1.3}
\begin{tabular}{c|ccc|ccc}
\hline
\multirow{2}{*}{\textbf{Fuzzing Framework}} & \multicolumn{3}{c|}{\textbf{Coverage, \%}} & \multicolumn{3}{c}{\textbf{Shannon Entropy (average)}} \\ \cline{2-7}
                                            & \textbf{Modbus/TCP} & \textbf{S7Comm} & \textbf{Ethernet/IP} & \textbf{Modbus/TCP} & \textbf{S7Comm} & \textbf{Ethernet/IP} \\ \hline
\textbf{MALF}                               & \textbf{91.23}         & \textbf{90.67}  & \textbf{89.48}       & \textbf{4.4}        & \textbf{4.1}    & \textbf{4.3}         \\
\textbf{NCMFuzzer}                          & 84.33              & 81.75           & 82.05               & 3.6                 & 3.3             & 3.7                  \\
\textbf{ChatAFL}                            & 74.05              & 60.5            & 60.1                & 3.6                 & 3.7             & 3.8                  \\
\textbf{Peach}                              & 57.85              & 46.67           & 59.75               & 3.2                 & 3.2             & 3.3                  \\ \hline
\end{tabular}
\label{tab:table3}
\end{table*}

\subsection{Ablation Study}
\begin{figure*}[!t]
\centering
\subfloat[]{\includegraphics[width=2in]{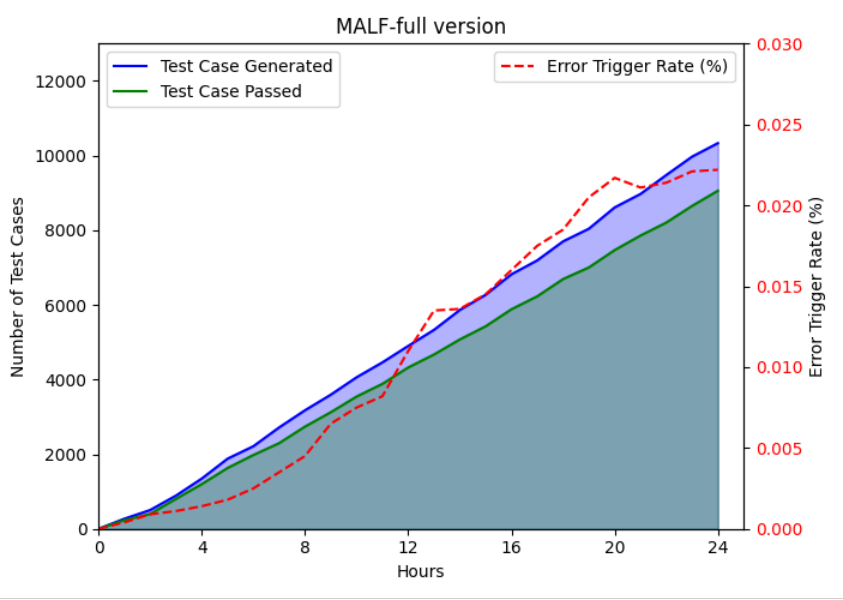}%
\label{fig_first_case}}
\hfil
\subfloat[]{\includegraphics[width=2in]{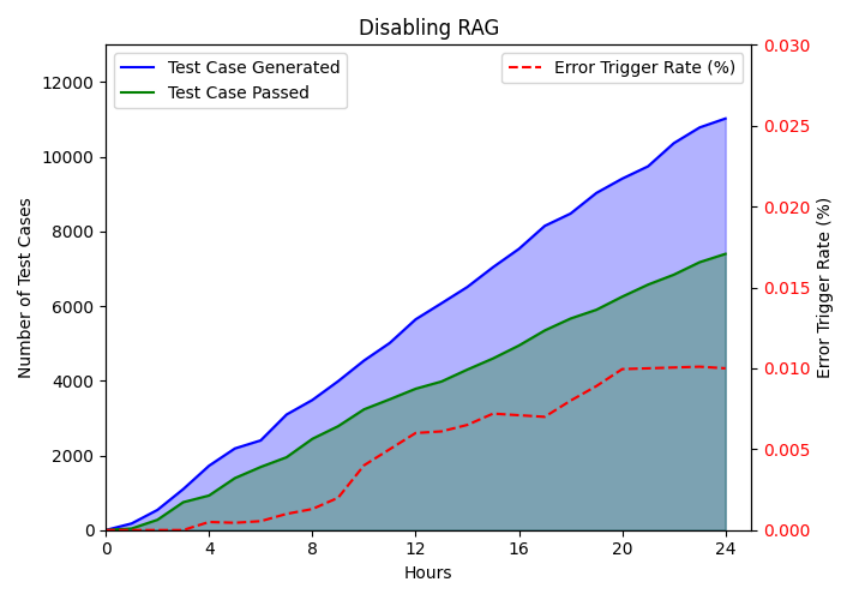}%
\label{fig_second_case}}
\hfil
\subfloat[]{\includegraphics[width=2in]{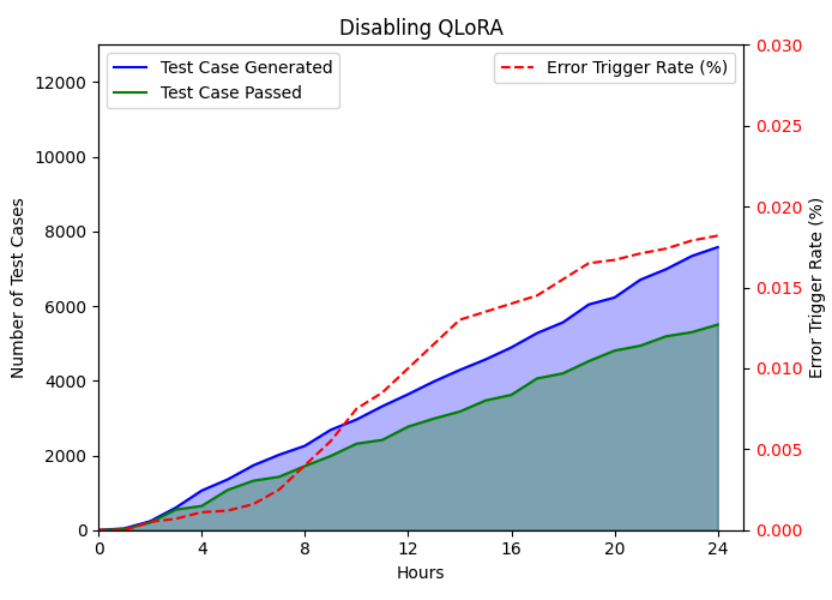}%
\label{fig_third_case}}
\caption{Test case pass rates and exception trigger rates over 24 hours for MALF and its ablated variants. (a) the original MALF framework, (b) MALF without RAG, and (c) MALF without QLoRA.}
\label{fig_12}
\end{figure*}

To assess the contributions of domain augmentation techniques, we conducted an ablation study on two key components of MALF: RAG and QLoRA. By selectively disabling these enhancements, we examined their impact on test case generation, TCPR, and ETN across ICPs.
Two experimental variants were tested. The first disabled RAG, removing access to domain-specific knowledge. The second disabled QLoRA, relying on the base LLM without ICS-specific fine-tuning while keeping RAG and multi-agent coordination intact. Results are shown in Fig. \ref{fig_12} and Table \ref{tab:table4}.

\subsubsection{Disabling RAG}
The total number of test cases generated shows a slight increase, as the system spends less time incorporating retrieved context and instead focuses on producing more general-purpose outputs. However, TCPR drops significantly, declining by around 15\%, due to the absence of protocol-specific data. This results in seeds and mutations that often deviate from vendor-defined constraints or involve uncommon function codes, leading to higher rejection rates during testing. Additionally, the ETN experiences a sharp decline, reflecting the framework's reduced ability to generate impactful test cases capable of uncovering vulnerabilities. Without real-world ICS references or CVE records, the test cases are less effective in pushing the SUT into critical error states.

\subsubsection{Disabling LoRA}
The LLM’s generation speed and interactive capabilities are limited due to the absence of memory-efficient fine-tuning that integrates ICS-specific knowledge.While multi-agent collaboration and RAG integration remain intact, the LLM operates on its original parameters, which are less calibrated for fuzzing tasks. This leads to a moderate performance decline, with TCPR reduced by 8–10\%. Though RAG provides domain knowledge, the base model struggles to integrate nuanced field relationships, diminishing its ability to generate precise protocol-compliant payloads. As the model’s reduced capability to assimilate retrieved context weakens its effectiveness in generating test cases that trigger exceptions.

These results underscore the critical roles of RAG and QLoRA in enhancing MALF’s domain-awareness and operational efficiency:

\romannumeral1.\ Context-Driven Fuzzing: RAG injects precise, up-to-date information about protocol variants, vulnerabilities, and domain best practices. Without it, the model’s ability to craft seeds and mutations reflecting realistic ICS quirks is substantially diminished.

\romannumeral2.\ Efficient Adaptation: LoRA superimposes domain-specific weight adjustments on top of the base LLM, letting the framework internalize crucial ICS knowledge without burdensome overhead. Its removal cripples the model’s capability to assimilate and apply specialized data gleaned from RAG, causing a marked drop in Fuzzing Effectiveness.

Beyond disabling RAG and QLoRA, we conceptually evaluate a variant in which Chain-of-Thought reasoning is omitted from all agents. CoT provides the ordered, fine-grained reasoning steps that let the Seed-Generation Agent verify cross-field invariants and enables the Test-Case Generation Agent to preserve length–value coherence during structural or semantic mutation. Without this scaffolding, seeds are likely to violate basic protocol constraints, causing malformed packets and sharply reducing the expected test-case pass rate. The resulting drop in syntactic validity also constrains mutation diversity, limiting exploration of deep protocol states. More critically, the absence of a transparent reasoning trail undermines inter-agent hand-offs: feedback cannot be mapped back to reproducible decision paths, leading to brittle orchestration and lower overall fault discovery. Hence, CoT is the logical glue that sustains disciplined multi-agent collaboration; its removal would erode the very advantages—high protocol compliance, adaptive mutation, and coordinated learning—that distinguish MALF from prior single-agent fuzzers.

By addressing these key components—RAG and LoRA, we shed light on the \textit{Q2}. The data clearly shows that context retrieval and targeted parameter training work in unison to heighten protocol compliance, escalate exception triggers, and enhance overall fuzzing efficiency. As such, we affirm the necessity of these enhancements for robust, high-coverage ICPs fuzzing.

\begin{table*}[!t]
\caption{Comparison of evaluation results for MALF Variant }
\centering
\renewcommand{\arraystretch}{1.3}
\begin{tabular}{lcccc}
\toprule
\textbf{MALF Variant} & \textbf{Number of Test Cases Generate} & \textbf{Number of Test Case Passed} & \textbf{TCPR} & \textbf{Error Trigger Rate} \\
\midrule
Full version & \textbf{10334} & \textbf{9059} & \textbf{87.66}\% & \textbf{0.022\%} \\
Disabling RAG & 11026 (106.69\%) & 7390 & 67.02\% & 0.011\% (50\%) \\
Disabling LoRA & 7580 (73.35\%) & 5605 & 73.95\% & 0.018\% (81.82\%) \\
\bottomrule
\end{tabular}
\label{tab:table4}
\end{table*}

\subsection{Real-World Industrial Testbed Vulnerability Discovery}

To further validate MALF, we deployed it in a cutting-edge cyber-physical testbed with multiple PLCs. Over three 24-hour fuzzing cycles, MALF autonomously identified several protocol-level flaws, including known CVEs and three zero-day vulnerabilities, which were missed by baseline tools. One of these zero-day issues, later confirmed as CNVD-2024-16009, was reported to the Chinese National Vulnerability Database (CNVD). Table \ref{tab:table5} summarizes the vulnerabilities discovered, followed by detailed descriptions of the zero-day findings.
While MALF reconfirmed several known CVEs, its most significant success lies in uncovering three zero-day vulnerabilities:

\begin{table*}[!t]
\caption{Overview of discovered vulnerabilities during three 24-hour fuzzing cycles on real PLCs.\label{tab:table5}}
\centering
\renewcommand\arraystretch{1.3}
\begin{tabularx}{\textwidth}{@{} XXXX@{} }
\toprule
\textbf{Device} & \textbf{Vulnerability Description} & \textbf{Security Issue} & \textbf{Reference} \\ 
\midrule
SIEMENS S7-300          & Denial of Service Vulnerability      & Device crashed                   & CVE-2016-3949           \\ 
                        & Input Validation Vulnerability                 & Device crashed            & CVE-2015-2177           \\ 
                        & Controller Password Protection Vulnerability     & Device access compromised          & CVE-2015-0004           \\ 
                        & Plaintext Unverified Protocol Vulnerability        & Data leakage or device exposure      & CVE-2015-0019           \\ 
                        & PLC Memory Read/Write Vulnerability            & Device memory corrupted          & CVE-2015-0015           \\ 
                        & I/O Module Stoppage              & Device functionality disrupted             & Under review
                       \\ \hline
SIEMENS S7-1200         & CPU Device CRLF Injection Vulnerability        & Device crashed          & CVE-2014-2909           \\ 
                        & CPU Device Cross-Site Scripting (XSS)         & Device compromised           & CVE-2014-2908           \\ 
                        & Network Server Cross-Site Scripting (XSS)        & Network data manipulated        & CVE-2012-3040           \\ 
                        & SIMATIC Denial of Service Vulnerability          & Device crashed        & CVE-2013-0700           \\ 
                        & S7-1200 Information Disclosure Vulnerability       & Data leakage       & CVE-2012-3037           \\ \hline
AB 1769-L30ER              & I/O Module Stoppage (Denial of Service)         & Device functionality disrupted         & CNVD-2020-62433         \\ 
                        & PLC Connection Denial of Service     & Device communication lost         & CNVD-2024-16009*        \\ \hline
AB 1766-L32BWA             & Buffer Overflow Vulnerability              & Device crashed              & CNVD-2018-00883         \\ \hline
Emerson IC200PWR102L            & Remote Terminal Unit Remote Code Execution       & Device compromised        & CNVD-2013-13377         \\ 
                        & Missing Authentication or Authorization       & Unauthorized access           & CVE-2022-2793           \\ 
                        & Plaintext Unverified Protocol Vulnerability   & Data leakage or device exposure           & Under review                       \\ \hline
MITSUBISHI FX5U-32MT            & Denial of Service Vulnerability        & Device crashed                  & CNVD-2019-33386         \\ 
\bottomrule
\end{tabularx}
\end{table*}

\begin{figure}[!h]
\centering
\includegraphics[width=3 in]{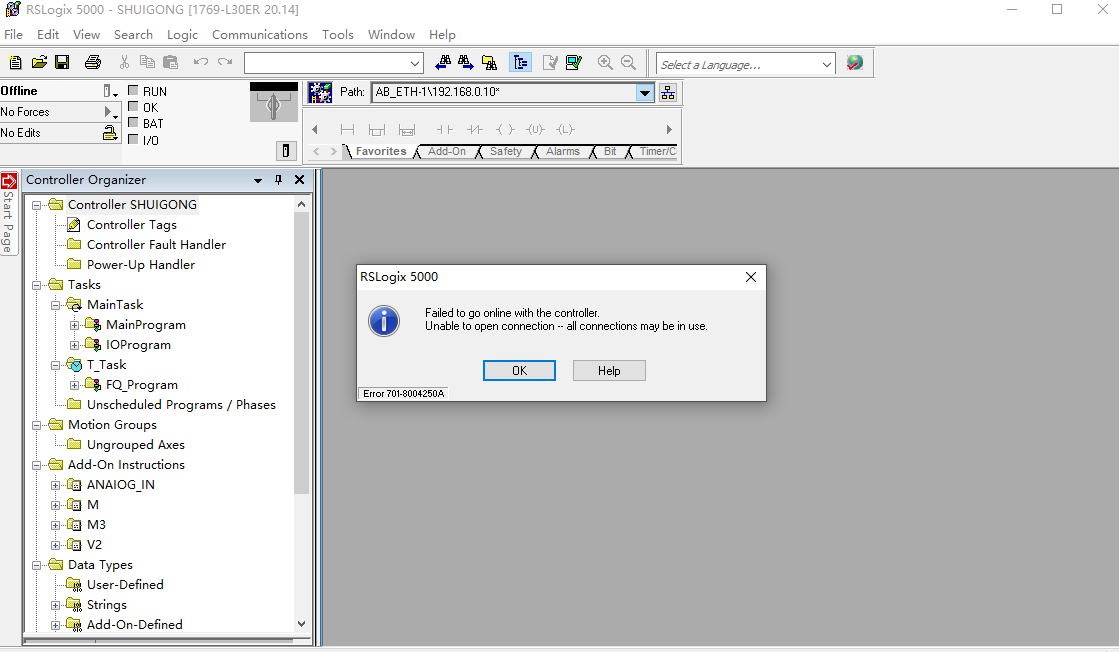}
\caption{Connection Failure Exception Diagram of CNVD-2024-16009.}
\label{fig_13}
\end{figure}

\begin{itemize}
    \item {1769-L30ER: PLC Connection Denial of Service (CNVD-2024-16009).
    
The Allen-Bradley 1769-L30ER controller establishes a session by responding to initial connection requests and awaiting termination signals when the session concludes. By omitting the explicit disconnection command in malformed “registration” packets, MALF discovered that the PLC fails to release allocated communication resources. Connection failures are monitored in RSLogix 5000 as shown in fig. \ref{fig_13} Consequently, an adversary can loop these incomplete sessions and gradually exhaust the PLC’s resource pool, effectively preventing legitimate users from accessing the controller. 
}
    \item {SIEMENS S7-300: I/O Module Stoppage.
    
Through targeted S7 protocol mutations—especially those exploring out-of-sequence I/O commands—MALF revealed a scenario where the PLC’s I/O modules unexpectedly halted when receiving malformed write instructions in rapid succession. Such repeated instructions circumvented typical handshake validations and left the module unresponsive, necessitating a manual reset. 
}
    \item {IC200PWR102L: Plaintext Unverified Protocol Vulnerability.
    
By comparing real-time responses with vendor-provided ICS documentation, MALF detected that critical data transfers occurred over an unverified, plaintext protocol, lacking even rudimentary authentication. This flaw opens the potential for eavesdropping and malicious command injection, rendering the device susceptible to remote manipulation. }
\end{itemize}

These real-world findings demonstrate the superior efficacy of MALF in uncovering critical defects, even within sophisticated ICS hardware. By automating the discovery of syntactic weaknesses and stateful protocol anomalies, the framework addresses a fundamental challenge in industrial cybersecurity: bridging the gap between theoretical vulnerabilities and actual exploit pathways in complex, resource-constrained PLC environments. We answer \textit{Q3} Beyond re-identifying historical CVEs, MALF’s ability to expose zero-day threats proves its tangible value for control system integrators and security practitioners alike.

As we expand the platform to support more protocols and incorporate richer ICS datasets, we expect even faster and more accurate vulnerability detection. By continuously refining its multi-agent orchestration and leveraging LLM-based reasoning with domain-specific intelligence, MALF is poised to transform industrial fuzz testing and fortify defenses against emerging threats in critical infrastructure.

\section{Discussion}
This work introduces the Multi-Agent LLM Fuzzing Framework (MALF), a groundbreaking approach to industrial protocol fuzz testing that combines multi-agent coordination, domain-specific knowledge retrieval, and adaptive mutation strategies. By addressing key challenges in fuzz testing, MALF makes significant contributions to both the research community and the industrial cybersecurity field.

\subsection{Challenges in Multi-Agent Coordination }
Seamless collaboration among agents is fundamental to MALF’s design. The framework addresses communication delays and task bottlenecks with asynchronous messaging via ZeroMQ, ensuring real-time data exchange. Task queues provide basic load balancing, but predictive workload distribution could further optimize agent performance. Fault tolerance is achieved through heartbeat signals and retry mechanisms, maintaining stability even during agent failures. Future enhancements, such as predictive fault detection and self-healing architectures, could further improve system resilience and scalability.

\subsection{Framework Performance Analysis}
MALF demonstrates outstanding performance in generating high-quality seeds and diverse mutations, achieving significant advancements in fuzz testing effectiveness and efficiency. In environments involving widely used industrial protocols such as Modbus/TCP, MALF achieves a TCPR as high as 92\%, a notable improvement over baseline methods. Furthermore, within a defined 24-hour testing cycle, MALF triggers significantly more exceptions, showcasing its superior capability to stress-test industrial control systems and identify critical vulnerabilities. The integration of RAG is pivotal to MALF’s success, enabling the framework to maintain protocol coverage exceeding 90\% while delivering Shannon Entropy values between 4.2 and 4.6 bits, reflecting its ability to generate highly diverse and protocol-compliant mutations. These attributes allow MALF to uncover latent vulnerabilities with fewer but more impactful test cases, a stark contrast to volume-driven fuzzers like Peach, which prioritize quantity over precision.

However, MALF's slightly lower throughput remains a trade-off, driven by its emphasis on generating high-quality, targeted test cases. To address this limitation, adopting parallelization strategies, such as multi-GPU processing and hardware acceleration, can enhance scalability while preserving its hallmark precision. By balancing throughput and impact, MALF establishes itself as a robust and adaptive framework, pushing the boundaries of fuzz testing in complex industrial environments.

\subsection{The Role of Large Language Models}
While the rise of LLMs has revolutionized cybersecurity, many existing approaches only use LLMs as auxiliary tools, relying heavily on manual intervention and extensive fine-tuning. These methods lack the scalability and automation required for comprehensive Industrial control protocol fuzzing. MALF overcomes these limitations by leveraging RAG and QLoRA to create a fully automated, end-to-end fuzzing pipeline. The domain-enhanced LLM is adapted to understand complex protocol semantics, dynamically generate diverse test cases, and respond effectively to real-time feedback. These capabilities surpass traditional rule-based fuzzing methods, enabling MALF to tackle intricate vulnerabilities in industrial protocols. 
Beyond fuzz testing, domain-enhanced LLMs hold potential for broader applications in industrial cybersecurity, including anomaly detection, threat intelligence analysis, and incident response, by leveraging their ability to process vast, unstructured data with contextual precision.

\subsection{Application Scenarios of MALF}
MALF stands out as a pioneering framework tailored to the unique demands of industrial control protocols, which differ fundamentally from the loosely structured network and text-based protocols targeted by conventional fuzzing tools. Protocols in industrial environments, with their strict formatting and stability requirements, necessitate advanced methods to ensure robust vulnerability mining without jeopardizing system operations. MALF is applicable to a wide range of industrial protocols, including Modbus/TCP, S7Comm, and Ethernet/IP, and has been successfully tested on real-world PLCs. Its modular architecture ensures adaptability to emerging technologies and complex environments In practical deployments, MALF has uncovered critical vulnerabilities, including zero-day exploits, while maintaining strict adherence to protocol constraints.
Through a series of experimental validations, MALF has demonstrated its capability to serve as a high-performance, fully automated, end-to-end, modular fuzzing framework for mainstream industrial control protocols. By seamlessly integrating precision, scalability, and adaptability, MALF sets a new standard for vulnerability discovery and robust protocol testing in safety-critical industrial environments.

\subsection{Future Directions}
MALF’s modular architecture opens pathways for significant enhancements. A plugin-based design could enable dynamic integration of high-speed RAG techniques, supporting additional protocols and advanced fuzzing methods. Expanding the framework to more complex environments, such as distributed control systems (DCS) or IoT ecosystems, could increase its applicability to modern industrial setups. Integration with real-time security operations centers (SOCs) would further extend its utility, enabling continuous vulnerability monitoring and proactive mitigation.

\section{Conclusion}
This work introduces a novel Multi-Agent LLM Fuzzing Framework (MALF), combining LLMs with multi-agent coordination to revolutionize ICP fuzz testing. By integrating RAG-driven domain-specific retrieval and QLoRA fine-tuning, MALF achieves precise seed generation, diverse mutations, and adaptive feedback optimization, surpassing baseline methods in test case pass rates, mutation diversity, and fuzzing efficiency. MALF exemplifies how LLMs can transform industrial cybersecurity, automating complex protocol analysis and uncovering hidden vulnerabilities. Its modular design enables scalability, adaptability, and integration into diverse industrial environments, laying a foundation for AI-driven innovations in network security. This research not only advances fuzz testing methodologies but also highlights the potential of LLMs to address evolving challenges in critical infrastructure protection, bridging the gap between traditional approaches and intelligent, automated solutions.

\section*{Patents}
An industrial control protocol fuzzing test benchmark platform. (Authorized Patent)

Inventors: Bowei Ning, Xuejun Zong, Bing Han, Kan He, Guogang Wang, Lian Lian, Yifei Sun and Hongyu Zheng.

Assignee: Zhigang Zhang.

Patent number: CN2024109705158.

Date of grant: 19 July, 2024.

Content: This invention presents a hybrid industrial control protocol fuzz testing benchmark platform, featuring virtualized communication, model loading, fuzz testing, database, performance evaluation, log monitoring modules, and physical industrial control networks. It scans and connects devices, loads models via APIs, sends test cases, manages test data and anomalies, and monitors device responses. Unified metrics enable comprehensive performance evaluation by comparing with specified tools. The modular, scalable design supports diverse models, standardizes fuzz testing evaluations, and accelerates the development and testing of high-performance industrial protocol fuzzing tools.

\section*{Data Availability}
Data will be made available on request.

\section*{Acknowledgments}
This work was supported in part by the Major Science and Technology Project of Liaoning Province (grant Nos. [2025]77(3)-1 and 2024JH1/11700049); the Applied Basic Research Program of Liaoning Province (grant No. 2025JH2/101300012); the Natural Science Foundation of Liaoning Province (grant No. 2023-MSLH-273); the Science and Technology Program of Liaoning Province (grant No. 2023JH1/10400082); and the Liaoning Provincial Science and Technology Innovation Platform Program (grant No. [2022]36).

\bibliographystyle{unsrt}  
\bibliography{references}

\end{document}